\documentclass[twocolumn]{aastex631} 
\graphicspath{{./}{figures/}}

\usepackage{amsmath}
\usepackage{bm}
\usepackage{color}
\usepackage{hyperref}
\makeatletter
\DeclareRobustCommand{\rvdots}{
  \vbox{
    \baselineskip4\p@\lineskiplimit\z@
    \kern-\p@
    \hbox{.}\hbox{.}\hbox{.}
  }}
\makeatother

\begin{document}
\newcommand{\todo}{\textcolor{red}}

\title{Investigating the sightline of a highly scattered FRB through a filamentary structure in the local Universe}

\author[0000-0002-6823-2073]{Kaitlyn Shin}
  \affiliation{MIT Kavli Institute for Astrophysics and Space Research, Massachusetts Institute of Technology, 77 Massachusetts Ave, Cambridge, MA 02139, USA}
  \affiliation{Department of Physics, Massachusetts Institute of Technology, 77 Massachusetts Ave, Cambridge, MA 02139, USA}
\author[0000-0002-4209-7408]{Calvin Leung}
  \altaffiliation{NASA Hubble Fellowship Program~(NHFP) Einstein Fellow.}
  \affiliation{Department of Astronomy, University of California, Berkeley, CA 94720, United States}
\author[0000-0003-3801-1496]{Sunil Simha}
  \affiliation{Center for Interdisciplinary Research in Astrophysics, Northwestern University, 1800 Sherman Ave., Evanston, IL 60201, USA}
  \affiliation{Department of Astronomy and Astrophysics, University of Chicago, William Eckhardt Research Center, 5640 South Ellis Avenue, Chicago, IL 60637}
\author[0000-0001-5908-3152]{Bridget~C. Andersen}
  \affiliation{Department of Physics, McGill University, 3600 rue University, Montr\'eal, QC H3A 2T8, Canada}
  \affiliation{Trottier Space Institute at McGill University, 3550 rue University, Montr\'eal, QC H3A 2A7, Canada}
\author[0000-0001-8384-5049]{Emmanuel Fonseca}
  \affiliation{Department of Physics and Astronomy, West Virginia University, PO Box 6315, Morgantown, WV 26506, USA }
  \affiliation{Center for Gravitational Waves and Cosmology, West Virginia University, Chestnut Ridge Research Building, Morgantown, WV 26505, USA}
\author[0000-0003-0510-0740]{Kenzie Nimmo}
  \affiliation{MIT Kavli Institute for Astrophysics and Space Research, Massachusetts Institute of Technology, 77 Massachusetts Ave, Cambridge, MA 02139, USA}
\author[0000-0002-3615-3514]{Mohit Bhardwaj}
  \affiliation{McWilliams Center for Cosmology, Department of Physics, Carnegie Mellon University, Pittsburgh, PA 15213, USA}
\author[0000-0002-1800-8233]{Charanjot Brar}
  \affiliation{NRC Herzberg Astronomy and Astrophysics, 5071 West Saanich Road, Victoria, BC V9E2E7, Canada}
\author[0000-0002-2878-1502]{Shami Chatterjee}
  \affiliation{Cornell Center for Astrophysics and Planetary Science, Cornell University, Ithaca, NY 14853, USA}
\author[0000-0001-6422-8125]{Amanda M.~Cook}
  \affiliation{David A.~Dunlap Department of Astronomy \& Astrophysics, University of Toronto, 50 St.~George Street, Toronto, ON M5S 3H4, Canada}
  \affiliation{Dunlap Institute for Astronomy \& Astrophysics, University of Toronto, 50 St.~George Street, Toronto, ON M5S 3H4, Canada}
\author[0000-0002-3382-9558]{B.~M.~Gaensler}
  \affiliation{Department of Astronomy and Astrophysics, University of California Santa Cruz, 1156 High Street, Santa Cruz, CA 95064, USA}
  \affiliation{Dunlap Institute for Astronomy \& Astrophysics, University of Toronto, 50 St.~George Street, Toronto, ON M5S 3H4, Canada}
  \affiliation{David A.~Dunlap Department of Astronomy \& Astrophysics, University of Toronto, 50 St.~George Street, Toronto, ON M5S 3H4, Canada}
\author[0000-0003-3457-4670]{Ronniy C Joseph}
  \affiliation{Department of Physics, McGill University, 3600 rue University, Montr\'eal, QC H3A 2T8, Canada}
  \affiliation{Trottier Space Institute at McGill University, 3550 rue University, Montr\'eal, QC H3A 2A7, Canada}
\author[0000-0003-3236-8769]{Dylan Jow}
  \affiliation{Kavli Institute for Particle Astrophysics and Cosmology, Stanford University, 452 Lomita Mall, Stanford, CA 94305, USA}
\author[0000-0003-4810-7803]{Jane Kaczmarek}
  \affiliation{CSIRO Space \& Astronomy, Parkes Observatory, P.O. Box 276, Parkes NSW 2870, Australia}
\author[0009-0007-5296-4046]{Lordrick Kahinga}
  \affiliation{Division of Physical and Biological Sciences, University of California Santa Cruz, Santa Cruz, CA 95064, USA}
\author[0000-0001-9345-0307]{Victoria M.~Kaspi}
  \affiliation{Department of Physics, McGill University, 3600 rue University, Montr\'eal, QC H3A 2T8, Canada}
  \affiliation{Trottier Space Institute at McGill University, 3550 rue University, Montr\'eal, QC H3A 2A7, Canada}
\author[0009-0008-6166-1095]{Bikash Kharel}
  \affiliation{Department of Physics and Astronomy, West Virginia University, PO Box 6315, Morgantown, WV 26506, USA }
  \affiliation{Center for Gravitational Waves and Cosmology, West Virginia University, Chestnut Ridge Research Building, Morgantown, WV 26505, USA}
\author[0000-0003-2116-3573]{Adam E.~Lanman}
  \affiliation{MIT Kavli Institute for Astrophysics and Space Research, Massachusetts Institute of Technology, 77 Massachusetts Ave, Cambridge, MA 02139, USA}
  \affiliation{Department of Physics, Massachusetts Institute of Technology, 77 Massachusetts Ave, Cambridge, MA 02139, USA}
\author[0000-0002-5857-4264]{Mattias Lazda}
  \affiliation{David A.~Dunlap Department of Astronomy \& Astrophysics, University of Toronto, 50 St.~George Street, Toronto, ON M5S 3H4, Canada}
  \affiliation{Dunlap Institute for Astronomy \& Astrophysics, University of Toronto, 50 St.~George Street, Toronto, ON M5S 3H4, Canada}
\author[0000-0002-7164-9507]{Robert A.~Main}
  \affiliation{Department of Physics, McGill University, 3600 rue University, Montr\'eal, QC H3A 2T8, Canada}
\author[0000-0003-4584-8841]{Lluis Mas-Ribas}
  \affiliation{Division of Physical and Biological Sciences, University of California Santa Cruz, Santa Cruz, CA 95064, USA}
\author[0000-0002-4279-6946]{Kiyoshi W.~Masui}
  \affiliation{MIT Kavli Institute for Astrophysics and Space Research, Massachusetts Institute of Technology, 77 Massachusetts Ave, Cambridge, MA 02139, USA}
  \affiliation{Department of Physics, Massachusetts Institute of Technology, 77 Massachusetts Ave, Cambridge, MA 02139, USA}
\author[0000-0002-0772-9326]{Juan Mena-Parra}
  \affiliation{Dunlap Institute for Astronomy \& Astrophysics, University of Toronto, 50 St.~George Street, Toronto, ON M5S 3H4, Canada}
  \affiliation{David A.~Dunlap Department of Astronomy \& Astrophysics, University of Toronto, 50 St.~George Street, Toronto, ON M5S 3H4, Canada}
\author[0000-0002-2551-7554]{Daniele Michilli}
  \affiliation{MIT Kavli Institute for Astrophysics and Space Research, Massachusetts Institute of Technology, 77 Massachusetts Ave, Cambridge, MA 02139, USA}
  \affiliation{Department of Physics, Massachusetts Institute of Technology, 77 Massachusetts Ave, Cambridge, MA 02139, USA}
\author[0000-0002-8897-1973]{Ayush Pandhi}
  \affiliation{David A.~Dunlap Department of Astronomy \& Astrophysics, University of Toronto, 50 St.~George Street, Toronto, ON M5S 3H4, Canada}
  \affiliation{Dunlap Institute for Astronomy \& Astrophysics, University of Toronto, 50 St.~George Street, Toronto, ON M5S 3H4, Canada}
\author[0009-0008-7264-1778]{Swarali Shivraj Patil}
  \affiliation{Department of Physics and Astronomy, West Virginia University, PO Box 6315, Morgantown, WV 26506, USA }
  \affiliation{Center for Gravitational Waves and Cosmology, West Virginia University, Chestnut Ridge Research Building, Morgantown, WV 26505, USA}
\author[0000-0002-8912-0732]{Aaron B.~Pearlman}
  \altaffiliation{Banting Fellow, McGill Space Institute~(MSI) Fellow, \\ and FRQNT Postdoctoral Fellow.}
  \affiliation{Department of Physics, McGill University, 3600 rue University, Montr\'eal, QC H3A 2T8, Canada}
  \affiliation{Trottier Space Institute at McGill University, 3550 rue University, Montr\'eal, QC H3A 2A7, Canada}
\author[0000-0002-4795-697X]{Ziggy Pleunis}
  \affiliation{Anton Pannekoek Institute for Astronomy, University of Amsterdam, Science Park 904, 1098 XH Amsterdam, The Netherlands}
  \affiliation{ASTRON, Netherlands Institute for Radio Astronomy, Oude Hoogeveensedijk 4, 7991 PD Dwingeloo, The Netherlands}
\author[0000-0002-7738-6875]{J.~Xavier Prochaska}
  \affiliation{Department of Astronomy and Astrophysics, University of California Santa Cruz, Santa Cruz, CA 95064, USA}
  \affiliation{Kavli Institute for the Physics and Mathematics of the Universe (Kavli IPMU), 5-1-5 Kashiwanoha, Kashiwa, 277-8583, Japan}
  \affiliation{Division of Science, National Astronomical Observatory of Japan, 2-21-1 Osawa, Mitaka, Tokyo 181-8588, Japan}
\author[0000-0001-7694-6650]{Masoud Rafiei-Ravandi}
  \affiliation{Department of Physics, McGill University, 3600 rue University, Montr\'eal, QC H3A 2T8, Canada}
\author[0000-0002-4623-5329]{Mawson W.~Sammons}
  \affiliation{Department of Physics, McGill University, 3600 rue University, Montr\'eal, QC H3A 2T8, Canada}
  \affiliation{Trottier Space Institute at McGill University, 3550 rue University, Montr\'eal, QC H3A 2A7, Canada}
\author[0000-0003-3154-3676]{Ketan R.~Sand}
  \affiliation{Department of Physics, McGill University, 3600 rue University, Montr\'eal, QC H3A 2T8, Canada}
  \affiliation{Trottier Space Institute at McGill University, 3550 rue University, Montr\'eal, QC H3A 2A7, Canada}
\author[0000-0002-2088-3125]{Kendrick Smith}
  \affiliation{Perimeter Institute for Theoretical Physics, 31 Caroline Street N, Waterloo, ON N25 2YL, Canada}
\author[0000-0001-9784-8670]{Ingrid Stairs}
  \affiliation{Department of Physics and Astronomy, University of British Columbia, 6224 Agricultural Road, Vancouver, BC V6T 1Z1 Canada}
\newcommand{\allacks}{
A.B.P. is a Banting Fellow, a McGill Space Institute~(MSI) Fellow, and a Fonds de Recherche du Quebec -- Nature et Technologies~(FRQNT) postdoctoral fellow.
A.M.C. is funded by an NSERC Doctoral Postgraduate Scholarship.
A.P. is funded by the NSERC Canada Graduate Scholarships -- Doctoral program.
B.\,C.\,A. is supported by an FRQNT Doctoral Research Award.
C. L. is supported by NASA through the NASA Hubble Fellowship grant HST-HF2-51536.001-A awarded by the Space Telescope Science Institute, which is operated by the Association of Universities for Research in Astronomy, Inc., under NASA contract NAS5-26555.
FRB research at UBC is supported by an NSERC Discovery Grant and by the Canadian Institute for Advanced Research.  The baseband recording system on CHIME/FRB is funded in part by a CFI John R. Evans Leaders Fund grant to I.H.S.
J.X.P., L.K, L.M., and S.S. acknowledge support from NSF grants AST-1911140, AST-1910471, and AST-2206490 as members of the Fast and Fortunate for FRB Follow-up team.
K.N. is an MIT Kavli Fellow.
K.R.S is supported by FRQNT doctoral resarch award.
K.S. is supported by the NSF Graduate Research Fellowship Program.
K.W.M. holds the Adam J. Burgasser Chair in Astrophysics and is supported by the Carl G. and Shirley Sontheimer Research Fund.
M.B is a McWilliams fellow and an International Astronomical Union Gruber fellow. M.B. also receives support from the McWilliams seed grant.
M.W.S. acknowledges support from the Trottier Space Institute Fellowship program.
S.S. is a Northwestern University and University of Chicago Brinson Postdoctoral Fellow.
V.M.K. holds the Lorne Trottier Chair in Astrophysics \& Cosmology, a Distinguished James McGill Professorship, and receives support from an NSERC Discovery grant (RGPIN 228738-13), from an R. Howard Webster Foundation Fellowship from CIFAR.
Z.P. is supported by an NWO Veni fellowship (VI.Veni.222.295).
}


\correspondingauthor{Kaitlyn Shin}
\email{kshin@mit.edu}

\begin{abstract}

Fast radio bursts (FRBs) are unique probes of extragalactic ionized baryonic structure as each signal, through its burst properties, holds information about the ionized matter it encounters along its sightline.
FRB~20200723B is a burst with a scattering timescale of $\tau_\mathrm{400\,MHz} >$1~second at 400~MHz and a dispersion measure of DM~$\sim$~244~pc~cm$^{-3}$.
Observed across the entire CHIME/FRB frequency band,
it is the single-component burst with the largest scattering timescale yet observed by CHIME/FRB.
The combination of its high scattering timescale and relatively low dispersion measure present an uncommon opportunity to use FRB~20200723B to explore the properties of the cosmic web it traversed.
With an $\sim$arcminute-scale localization region, we find the most likely host galaxy is NGC~4602 (with PATH probability $P(O|x)=0.985$), which resides $\sim$30~Mpc away within a sheet filamentary structure on the outskirts of the Virgo Cluster.
We place an upper limit on the average free electron density of this filamentary structure of $\langle n_e \rangle~<~4.6^{+9.6}_{-2.0} \times 10^{-5}$~cm$^{-3}$,
broadly consistent with expectations from cosmological simulations.
We investigate whether the source of scattering lies within the same galaxy as the FRB, or at a farther distance from an intervening structure along the line of sight.
Comparing with Milky Way pulsar observations, we suggest the scattering may originate from within the host galaxy of FRB~20200723B.

\end{abstract}


\section{Introduction}
\label{sec:intro}

On large scales, ionized matter is found in galaxies, galaxy clusters, and a ``cosmic web'' of filamentary structures including knots, sheets, and filaments \citep{bond+1996}.
Simulations have predicted that this cosmic web can contain around half of the baryons in the Universe \citep{cen_ostriker_2006}.
This ionized matter is often so diffuse that it is challenging to observe even in the local Universe, requiring a patchwork of techniques --- such as UV and X-ray absorption spectroscopy, and X-ray emission observations --- to probe its emission \citep[e.g.,][]{eckert+2015,tejos+2016, nicastro+2018,deGraaff+2019,tanimura+2019,tanimura+2021}.
As largely extragalactic flashes of radio emission, fast radio bursts \citep[FRBs;][]{lorimer+2007} provide a way to probe both the column density and density variations of diffuse matter along their sightlines using their observed properties.
In particular, the observed dispersion measure (DM) of an FRB quantifies the integral column density of free electrons along the line of sight, and the observed temporal pulse broadening is a result of scattering, which is due to multi-path propagation through inhomogeneous media;
the latter property is characterized by a scattering timescale at an observed frequency.

A few earlier studies have demonstrated the power of FRBs to illuminate the physical nature of baryonic structures in our Universe.
\citet{prochaska+2019_halo} report on FRB~20181112A, which has an arcsecond localization that intersects the halo of a foreground galaxy, to constrain the diffusiveness of its halo gas.
\citet{connor+2020} use FRB~20191108A, which intersects the M31 and M33 halos, to constrain their shared plasma environment properties.
\citet{faber+2024_scattered} conduct a sightline analysis of FRB~20221219A to propose the burst could be scattered through the circumgalactic medium (CGM) of an intervening galaxy.
\citet{connor+2024_igm} use a sample of cosmological FRBs localized to host galaxies to measure the fraction of baryons in the cosmic web.
All of these studies primarily use measurements of dispersion and scattering to constrain ionized baryonic structure difficult to study via other methods.

Here, we report on the CHIME/FRB discovery of FRB~20200723B, a local Universe FRB that resides $\approx 4$ virial radii ($R_\mathrm{vir}$) away from the center of the Virgo Cluster in the ``W-M sheet'',
a filamentary structure on the outskirts of the Virgo Cluster \citep{kim+2016_virgocluster}.
We use the DM of FRB~20200723B to place constraints on the gas properties of the filamentary structure its host galaxy resides in.
Furthermore, we combine the DM with the measured scattering timescale to dissect the line of sight and investigate where the majority of the scatter broadening originates from.

In Section~\ref{sec:discovery}, we present the CHIME/FRB discovery of FRB~20200723B and its burst properties.
In Section~\ref{sec:localization}, we present the localization region for FRB~20200723B as well as its most probable host galaxy.
In Section~\ref{sec:line_of_sight_DM}, we discuss the DM budget for the FRB~20200723B sightline, noting its path through the W-M sheet.
In Section~\ref{sec:scattering}, we discuss the high scattering timescale observed for FRB~20200723B, unusual among bursts observed by CHIME/FRB.

\section{FRB~20200723B}
\label{sec:discovery}

\subsection{Discovery with CHIME/FRB}
\label{subsec:chimefrb}
The Canadian Hydrogen Intensity Mapping Experiment (CHIME) is a telescope located at the Dominion Radio Astronomical Observatory (DRAO) near Pencticton, British Columbia, Canada.
CHIME has 1024 dual-polarization antennas across the focal lines of four cylindrical paraboloid reflectors \citep{chime_sys_overview}, and it surveys the entire sky roughly daily at declination~$>$$-$11$^\circ$ at 400--800~MHz.
The CHIME/FRB project is a digital backend on the CHIME telescope. 
CHIME/FRB discovers $\sim$2-3 FRBs per day \citep{catalog1};
such a large discovery rate is enabled by the large field of view and high sensitivity of the CHIME telescope.

In order to identify these FRBs, 1024 fast Fourier transform (FFT) beams are formed and searched for FRBs in real time using their total intensity data,
which are at a downsampled time resolution of $\sim$1~ms \citep{ng+2017_fft, chimefrb_sys_overview}.
During the real-time search, Stokes-$I$ intensity data are searched for dispersed signals while the channelized voltage data -- baseband data -- are stored in a ring memory buffer.
All candidate bursts with detection signal-to-noise ratios S/N~$>$~8 have intensity data written to disk, but only brighter candidate bursts, e.g., typically with S/N~$>$~12, have their baseband data triggered and saved to disk.
The baseband data have a time resolution of 2.56~$\mu$s and typically span $\sim$100~ms around the signal of interest, depending on the time of arrival (TOA) and DM values --- as well as uncertainty on the DM value --- identified by the real-time detection pipeline.
Further details on how the baseband data are captured are provided by \citet{michilli+2021_baseband_pipeline} and \citet{basecat1}.

\begin{figure*}[htbp]
    \centering
    \includegraphics[width=0.9 \textwidth]{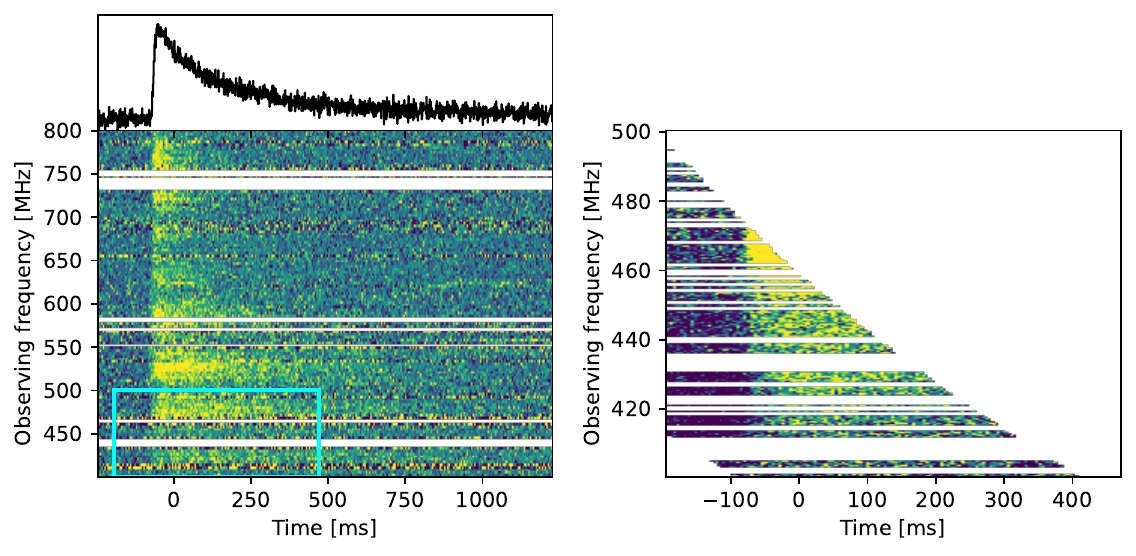}
    \caption{
        Dynamic spectra of FRB~20200723B as seen with intensity data on the left at 0.983~ms time resolution, and baseband data on the right downsampled to 5.12~ms time resolution from its recorded 2.56~$\mu$s time resolution.
        The intensity dynamic spectrum is plotted with the frequency channels downsampled by a factor of 8 at a resolution of 3~MHz (showing 128 frequency subbands), over the full 400~MHz observing bandwidth;
        the baseband dynamic spectrum is plotted at the recorded frequency resolution of 390~kHz, with only the bottom 100~MHz of the data capture shown.
        The cyan box overlaid on the intensity data subplot corresponds to the extent of the baseband data subplot.
        Both plots have bands of radio frequency interference (RFI) and missing data masked out.
        The time axes are referenced to the best-fit TOA from \texttt{fitburst}.
        Due to incorrect real-time detection parameters (e.g., DM and TOA), the baseband system only captured part of the burst.
        It is clear that the intensity data thus provide a much richer view of the burst properties of this FRB.
    }
    \label{fig:waterfall_comparison}
\end{figure*}

FRB~20200723B had a real-time detection S/N of $\approx$~40.1, resulting in a baseband dump being successfully triggered.
However, only a portion of the burst was caught in baseband.
Upon further inspection, this partial capture was likely due to the inaccurate identification of the true DM by the real-time pipeline because of the temporal scatter-broadening,
combined with the long duration of the pulse and the shorter timespan of baseband data dump compared to intensity data.
Nonetheless, despite us having captured only $\sim$100~MHz of signal at the bottom of the CHIME/FRB frequency band, the burst was bright enough to allow for an $\sim$arcminute scale offline localization and robust probabilistic association with a host galaxy (Section~\ref{sec:localization}).
Dynamic spectra of FRB~20200723B using the intensity data and the baseband data can be seen in Figure~\ref{fig:waterfall_comparison}.
The wider usable bandwidth and timespan of the lower time resolution intensity data offer much more reliable measurements of the burst properties of FRB~20200723B.
Thus, in the following sub-section, we focus on burst properties as observed using intensity data.

\subsection{Burst properties}
\label{subsec:burst_props}

We used a least-squares fitting framework called \texttt{fitburst} \citep{fonseca+2023_fitburst} to estimate the burst properties of FRB~20200723B.
The \texttt{fitburst} model fits a pulse broadening function for an intrinsically Gaussian and dispersed pulse;
the model parameters consist of the DM, burst TOA, signal amplitude, temporal pulse width, spectral index, spectral running, and scattering timescale $\tau$ at a reference frequency, in this case 400.1953125~MHz, the center of the lowest CHIME observing frequency channel \citep{catalog1}.
Most \texttt{fitburst} measurements for CHIME/FRB bursts typically fix the scattering index assuming $\tau \propto \nu^{-4}$ \citep[e.g.,][]{lorimerkramer2012_handbook, catalog1}.
As FRB~20200723B is bright and broadband, however, we also use \texttt{fitburst} to fit simultaneously for the scattering index.

\begin{figure}[htbp]
    \centering
    \includegraphics[width=1.0\columnwidth]{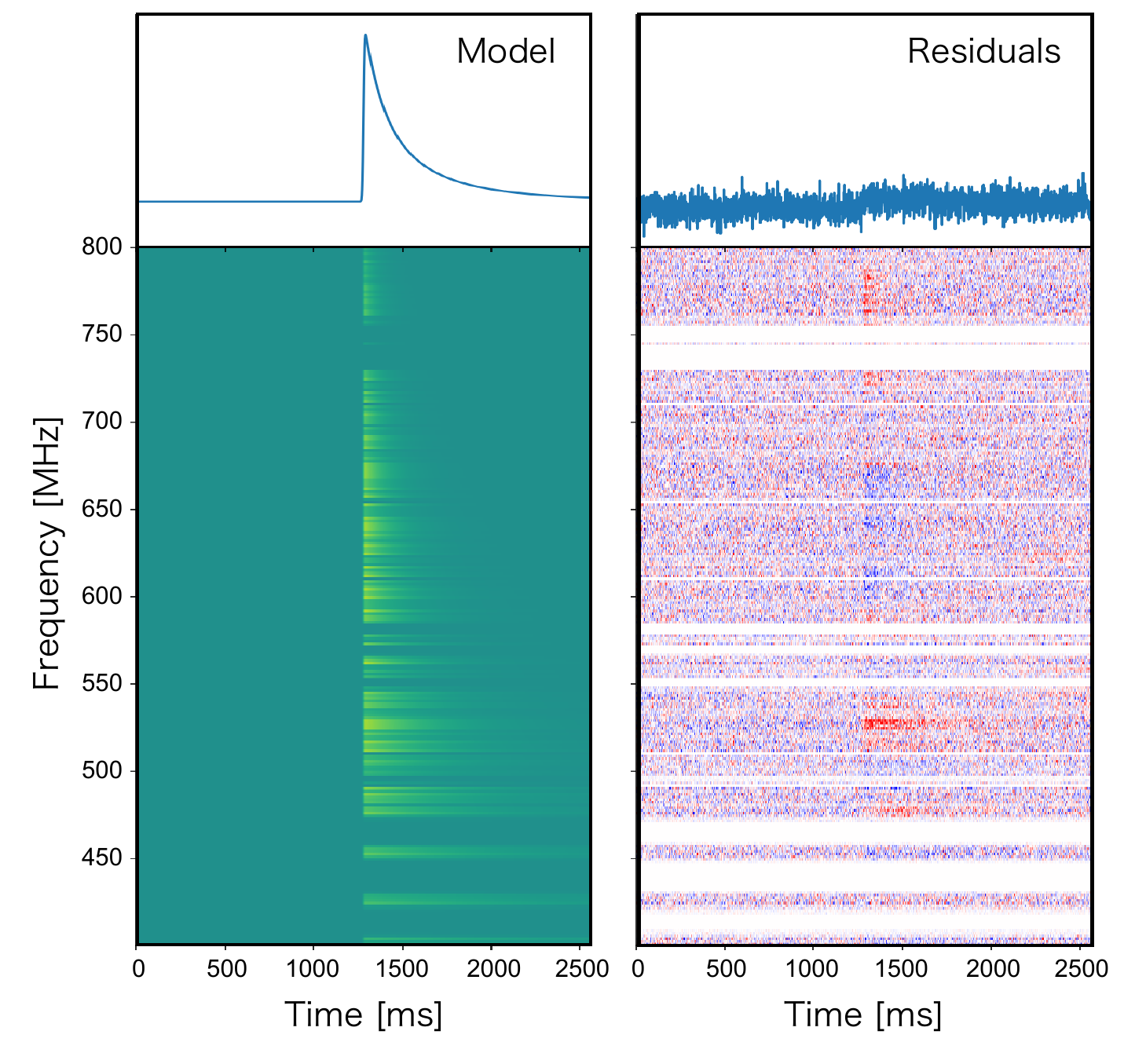}
    \caption{
        The best-fit \texttt{fitburst} model dynamic spectrum for FRB~20200723B (left) alongside the model residuals (right).
        The model is plot with 256 frequency channels; bands of RFI from the intensity data are masked out in both panels.
    }
    \label{fig:fitburst_104235227}
\end{figure}

For FRB~20200723B, \texttt{fitburst} measures DM~$=~243.99~\pm~0.06$~pc~cm$^{-3}$,
a scattering index $\alpha = -4.298 \pm 0.065$
and a scattering timescale $\tau_\mathrm{400\,MHz} = 1.140 \pm 0.026$~s at the reference frequency.
The \texttt{fitburst} fit to the intensity data of FRB~20200723B, as well as its model residuals, are shown in Figure~\ref{fig:fitburst_104235227}.
This scattering index is closer to $-4.4$, the value for a diffractive Kolmogorov scattering model, though it is still in between the values commonly adopted for scattering indexes ($-4.0$ to $-4.4$).
In practice, many pulsars show a flatter index than expected from diffractive Kolmogorov turbulence
\citep[i.e., $\alpha > -4.4$;][]{rickett+2009,krishnakumar+2017}.
If we fix $\alpha = -4$, as was done by \citet{catalog1}, we obtain with \texttt{fitburst} a 
DM~$=~244.05~\pm~0.06$~pc~cm$^{-3}$ and a scattering timescale $\tau_\mathrm{400\,MHz} = 1.025 \pm 0.009$~s at the reference frequency.
This model fit is marginally worse than the model fit obtained with the scattering index free \citep[using the F-test for model selection, e.g.,][]{catalog1}, but we report these numbers for comparison with the larger CHIME/FRB sample.
For all analyses through the rest of this paper, we use the values obtained from the \texttt{fitburst} fit where we fit for the scattering index as well.

This scattering timescale $\tau_\mathrm{400\,MHz} = 1.140 \pm 0.026$~s places FRB 20200723B among the most scattered FRBs discovered.
Of all the bursts yet reported by CHIME/FRB, only FRB~20191221A --- the FRB source demonstrating sub-second periodicity, with overlapping burst components --- has a larger scattering timescale, reported as $\tau_\mathrm{600\,MHz} = 340 \pm 10$~ms at 600~MHz \citep{chimefrb_subsecperiodicity}.
Assuming\footnote{In the analysis by \citet{chimefrb_subsecperiodicity}, they fit the integrated profile assumed at 600 MHz; they therefore did not fit for a frequency evolution.} $\tau \propto \nu^{-4}$, this corresponds to a scattering timescale of $\tau_\mathrm{400\,MHz} = 1.721 \pm 0.051$~s at the reference frequency of the bottom of the CHIME observing band.

\subsubsection{Flux/fluence}
\label{subsubsec:burst_props_flux}

Estimating fluxes or fluences from CHIME/FRB intensity data is challenging, in large part due to header localization limitations and the complex CHIME/FRB beam model \citep{andersen+2023_chimefrb_flux}.
(The header localization is obtained by fitting the detection S/N per formed beam to the frequency-dependent beam model.)
As such, the resulting flux and fluence measurements are best interpreted as lower limits, as was the case for the values reported by \citet{catalog1}.
Using the automated flux calibration software pipeline, we estimate the lower limit on the peak flux of the burst to be $\gtrsim 4.8 \pm 1.4$~Jy, and on the fluence of the burst to be $ \gtrsim 820 \pm 213$~Jy~ms.
All flux and fluence values are band-averaged over the CHIME 400--800~MHz observing frequencies.

This pipeline assumes that the burst was detected along the meridian of the primary beam, a simplifying assumption necessitated by the limited precision of the header localizations;
however, due to the captured baseband data for this burst, we are able to get a localization more accurate than the header localization (Section~\ref{sec:localization}).
Thus, in theory, our knowledge of the off-meridian position of the FRB should allow for us to correct for the response of the CHIME/FRB primary and synthesized beam pattern.
In practice, however, this burst is detected at a low declination, where the formed FFT beams exhibit more extreme ``clamping'' structure in the beam shape \citep{andersen+2023_chimefrb_flux}.
Thus, while this burst is detected just within the full-width half-max (FWHM) of the primary beam, we find that our formed beam sensitivity to this position is extremely low, especially at frequencies greater than $\sim$600~MHz.
We therefore do not report an intensity flux/fluence estimate using the baseband localization.

\subsubsection{Scintillation}
\label{subsubsec:burst_props_scint}

Scintillation is a frequency-dependent variation of intensity due to interference between scattered paths.
The scintillation bandwidth, or the decorrelation bandwidth, goes as $\Delta \nu_\mathrm{d} \propto \nu^{-\alpha}$.
While a single screen can cause both observable scatter-broadening (which goes as $\tau \propto \nu^\alpha$) and scintillation (which can be related to scattering with $\Delta \nu_\mathrm{d} \simeq (2 \pi \tau)^{-1}$),
it is also possible for scattering and scintillation to be contributed by separate screens \citep[e.g.,][]{masui+2015,ocker+2022_r1twin_largeDMh,sammons+2023_twoscreen,nimmo+2024_scintillation}.
A measurement of scintillation from a different screen than the screen that contributes the scattering can provide possible constraints on the locations of the scattering media along the line of sight.
At 1~GHz, the NE2001 Galactic electron density model \citep{ne2001} predicts a scatter-broadening of $\tau_\mathrm{1\,GHz} = 1.2 \times 10^{-4}$~ms,
and the expectation for the scintillation bandwidth is $\Delta \nu_\mathrm{d} = 1.6$~MHz.
Assuming $\alpha = -4.3$, these translate to predictions at 400~MHz of $\tau_\mathrm{400\,MHz} = 6.0 \times 10^{-3}$~ms and $\Delta \nu_\mathrm{d} = 0.03$~MHz.
As the observed scattering timescale for FRB~20200723B is greatly in excess of Galactic expectations, it is unlikely for the Milky Way to be the dominant source of observed scattering.

With 16,384 intensity channels across the 400~MHz-wide observing bandwidth of CHIME/FRB, we would be sensitive to spectral ``scintles'' on scales larger than $\gtrsim$48.8~kHz.
The NE2001 expectation for the scintillation bandwidth at 600~MHz is $\Delta \nu_\mathrm{d} \approx 174$~kHz, a scale we should be able to probe with our intensity data.
However, any scintillation scale associated with the same extragalactic screen that contributed the large scattering would have a decorrelation bandwidth $\Delta \nu_\mathrm{d} < 10^{-3}$~kHz and thus not be resolvable by our intensity data.
Importantly, there are artificial frequency variations introduced due to telescope systematics: the original channelization of the baseband data, 400~MHz over 1024 frequency channels, introduces a ``scalloping'' artifact at $\sim$0.38~MHz, and there is also an instrumental ``ripple'' at $\sim$30~MHz due to reflections off the feed structure within the CHIME telescope
\citep{chime_sys_overview}.
When searching the intensity data, no frequency variation scales were found other than the expected artificial variation scales.
The NE2001 prediction for scintillation may have been undetected because another screen resolved out the scintillation, or because it was washed out by the instrumental scalloping and correction.
It is also possible that a modulation index --- a measure of the flux variation caused by scintillation over the bandwidth --- that is $< 1$ could make detection more difficult.
Additionally, any intrinsic scintillation with frequency scales around the artificial variation scales, or with frequency scales at finer resolution than 48.8~kHz, cannot be ruled out.
It is possible to upchannelize the baseband data to explore scintillation at a finer scale \citep[e.g.,][]{schoen2021scintillation,nimmo+2024_scintillation}.
However, the limited bandwidth of the baseband data for FRB~20200723B limits this analysis.

\subsubsection{Polarimetry}
\label{subsubsec:burst_props_pol}

FRBs are often highly linearly polarized \citep[e.g.,][]{masui+2015, pandhi+2024_basecat_pol}, and linearly polarized radio waves undergo Faraday rotation as they pass through magneto-ionic material.
The Faraday rotation measure (RM) is related to the electron number density, $n_e$, and the line-of-sight ($l$) component of the magnetic field, $B_\parallel$, by RM $= 0.8 \int n_e(l) B_\parallel(l) dl$~rad~m$^{-2}$.
Previous observations of repeating FRBs have shown evidence for a correlation between high RM and strong scattering local to the burst \citep[e.g.,][]{feng2022frequency,ocker+2023_scatteringvariability, annathomas+2023} as well as the time-variability of the RM itself \citep{mckinven+2023_R3}, suggesting that the RM in at least some sources probes the magneto-ionic environment local to the FRB.
Motivated by these results, we measure the RM of FRB~20200723B.

We detect a strong, uncalibrated signal in the Stokes~$U$ profile of FRB~20200723B.
In large part due to the limited bandwidth of the baseband data capture, we are unable to calibrate the Stokes profiles.
For the same reason, the Stokes $QU$-fitting method of determining the RM fails.
This method is more robust against leakage introduced by path length differences in the CHIME telescope system \citep{mckinven+2021_polarization_pipeline}.
As such, robust polarization fraction estimates --- corrected for leakage --- cannot be obtained.
However, we are still able to constrain the RM using the baseband data using the RM synthesis method \citep{burn_1966, brentjens_deBruyn_2005_RM_synth} as implemented for CHIME/FRB by \citet{mckinven+2021_polarization_pipeline}.
This method does not require absolute calibration of the Stokes profiles.
The resulting Faraday dispersion function appears to be a convolution of instrumental effects and an astrophysical signal.
Thus, over the bottom $\sim$100~MHz bandwidth of the baseband data, we estimate the RM of FRB~20200723B to be $|$RM$|$~$< 60$~rad~m$^{-2}$.
(Using the all-sky map of the Milky Way RM contribution constructed by \citet{hutschenreuter+2022_MW_RM_map}, the Galactic foreground RM expectation is $-5 \pm -6$~rad~m$^{-2}$.)
This RM value for FRB~20200723B is quite modest, and indicates that the source probably does not originate from an extremely dense and/or strongly magnetized circumburst environment.

\section{Burst Localization and Host Galaxy}
\label{sec:localization}

\begin{figure}[htbp]
    \centering
    \includegraphics[width=1.0\columnwidth]{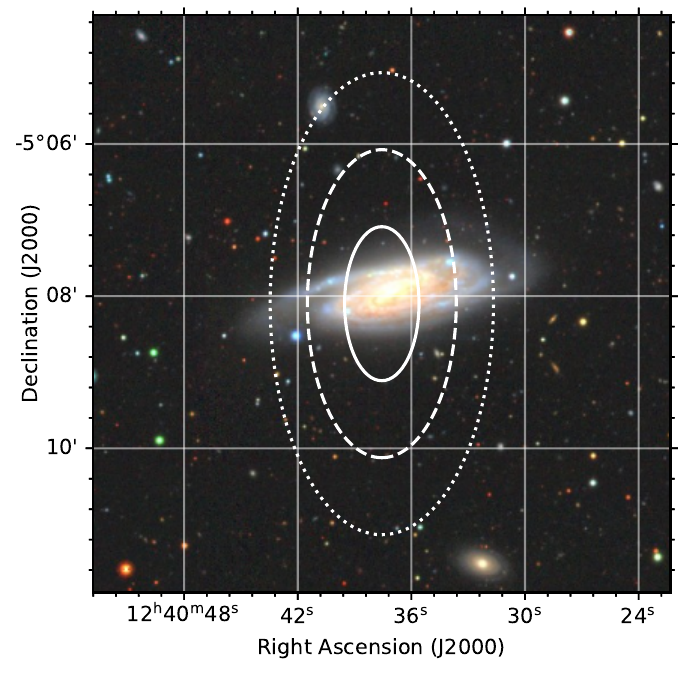}
    \caption{
        The 1$\sigma$ (solid ellipse), 2$\sigma$ (dashed ellipse), and 3$\sigma$ (dotted ellipse) baseband localization regions for FRB~20200723B, plotted on top of archival DECaLS imaging in the \textit{grz} bands.
        The central galaxy, and the putative host of FRB~20200723B, is NGC~4602 at $z=0.0085$.
    }
    \label{fig:bb_overlaid}
\end{figure}

Using the baseband localization pipeline \citep{michilli+2021_baseband_pipeline},
the resulting localization for FRB~20200723B is RA~=~12h40m38(2)s, Dec~=~$-$05$^{\circ}$08(1)$'$06$''$ (J2000), with the errors quoted as 1$\sigma$~uncertainties.
Figure~\ref{fig:bb_overlaid} shows this localization region overlaid on archival imaging from the Dark Energy Camera Legacy Survey \citep[DECaLS;][]{dey+2019_decals_overview}.
Clearly evident in the localization region is the bright putative host galaxy, NGC~4602, a Seyfert spiral galaxy \citep{tommasin+2010}.
This galaxy is included in the H\,\textsc{i} Parkes All-Sky Survey (HIPASS) catalog \citep{meyer+2004_hipass};
the reported neutral hydrogen radial velocity measurements imply a redshift of $z=0.0085$.
This redshift is consistent with what has been found with optical and mid-infrared spectroscopy \citep{tommasin+2008,feltre+2023}.
The redshift-inferred distance, using \textit{Planck} cosmological parameters \citep{planck2018vi}, is $\sim$37.59~Mpc.
This galaxy also has a redshift-independent distance as derived by the Tully-Fisher relation \citep{tully_fisher_relation_1977}.
Converting from the reported best-fit distance modulus, NGC~4602 is $32.49^{+2.87}_{-2.64}$~Mpc away, with an inclination angle of $i = 73 \pm 3^\circ$ \citep{kourkchi+2020_cosmicflows}.
Throughout this paper, we will use the redshift-independent distance when referring to NGC~4602, as it is more robust against peculiar velocity effects when inferring distance \citep[e.g.,][]{steer+2017}.
This host galaxy distance of $\approx$32.5~Mpc places FRB~20200723B among the closest reported FRBs;
other notably nearby FRBs include FRB~20200120E at $\approx$3.6~Mpc \citep{bhardwaj+2021_m81, kirsten+2022_m81} and
FRB~20181030A at $\approx$20~Mpc \citep{bhardwaj+2021_zmax}.
A further analysis and discussion of the host galaxy properties of NGC~4602 will be presented as part of a larger sample by Andersen et al. (in prep.).

We can also obtain a probabilistic maximum redshift for an FRB given its observed DM, i.e., $P(z|\mathrm{DM})$.
While both \citet{shin+2023} and \citet{james2023_chimezdm} model $P(z|\mathrm{DM})$ for CHIME/FRB based on observations from the first CHIME/FRB catalog, \citet{shin+2023} explicitly exclude highly scattered FRBs in their modeling.
Thus, we use the work by \citet{james2023_chimezdm}, who include highly scattered FRBs in their modeling,
and provide a joint probability distribution of redshift and extragalactic DM, $P(z, \mathrm{DM}_\mathrm{EG}).$\footnote{Available in the \texttt{FRBs} repository at \url{https://github.com/FRBs/zdm/blob/99aed43/docs/nb/CHIME_pzDM.ipynb}.}
We can go from the joint probability distribution to the conditional probability distribution via $P(z|\mathrm{DM}_\mathrm{EG}) = P(z, \mathrm{DM}_\mathrm{EG}) / P(\mathrm{DM}_\mathrm{EG})$.
Subtracting the NE2001 \citep{ne2001} model for the Galactic DM contribution to FRB~20200723B, we obtain a DM$_\mathrm{EG}$ contribution of $\approx$210~pc~cm$^{-3}$.
The 95\% upper limit of $P(z|\mathrm{DM}_\mathrm{EG}=210$~pc~cm$^{-3})$ is $z \approx 0.32$.
The redshift of NGC~4602, at $z=0.0085$, is comfortably within the range of redshifts one would expect based on its extragalactic DM, i.e., consistent with $z \leq 0.32$.

With NGC~4602 located so prominently within the 1$\sigma$ baseband localization region of FRB~20200723B at a redshift allowed by the observed DM, it is qualitatively unlikely for any other galaxy to be the host galaxy for this FRB.
Nevertheless, this association of FRB~20200723B with NGC~4602 can be assessed with the Probabilistic Association of Transients to their Hosts (PATH) method \citep{aggarwal+2021_PATH}.
PATH is a Bayesian framework which assigns association probabilities to each galaxy in a provided flux-limited catalog.
We use DECaLS as our galaxy catalog, as it is the deepest survey covering the localization region of FRB~20200723B.
Based on sandboxing simulations with host galaxy associations using PATH and CHIME/FRB baseband localization regions (Andersen et al. in prep.), we adopt an unseen prior\footnote{This unseen prior, $P(U)$ corresponds to the probability that the true host galaxy of the FRB is unseen within the adopted galaxy survey.
Typical values in the literature adopt $P(U)=0.1$ when using DECaLS as the galaxy survey \citep[e.g.,][]{ibik+2024}, but Andersen et al. (in prep.) found that a higher unseen prior probability was more representative for associations using CHIME/FRB localization regions.} $P(U)=0.15$.
The resulting posterior association probability for NGC~4602 is $P(O|x)=0.985$, a probabilistically robust host galaxy association.
No other galaxy detected by DECaLS has a posterior association probability of $P(O|x)>0.005$.

\section{Line of sight contributions to the DM}
\label{sec:line_of_sight_DM}

In this section, we quantify possible contributions to the DM of FRB~20200723B from media along its line of sight, noting that there is also a local Universe filamentary structure along the sightline of this burst.

\subsection{The W--M sheet}
\label{subsec:wm_sheet}

The nearest galaxy cluster to NGC~4602, both in 3D and projected on the sky, is the Virgo Cluster, which has a virial radius $R_\mathrm{vir} = 1.08$~Mpc \citep{urban+2011_virgocluster}.
\citet{reiprich+2013} define the outskirts of clusters to be out to $3 R_\mathrm{vir}$.
As the impact parameter between NGC~4602 and the Virgo Cluster is $b_\perp \sim 4.4$~Mpc $> 3 R_\mathrm{vir}$, there is likely a negligible amount of the intra-cluster medium (ICM) by NGC~4602.
Thus, we expect any potential ICM contribution to the total observed DM of FRB~20200723B to be negligible.
However, NGC~4602 appears to be part of a filamentary structure associated with the outskirts of the Virgo Cluster; this structure is called the ``W-M sheet'' \citep{kim+2016_virgocluster}.
Although the exact data comprising the discovered W-M sheet structure were not made public, because NGC~4602 is at a distance $< 40$~Mpc from Earth, we expect the galaxies in the W-M sheet to be encompassed by the latest NED Local Volume Sample \citep[NED-LVS 2021-09-22v2;][]{cook+2023_nedLVS}, which contains galaxies with distances out to 1000~Mpc.
To confirm that NGC~4602 is \textit{within} the W-M sheet, we query the NED-LVS for galaxies with projected positions within $\pm$2.5~degrees of the center of NGC~4602 and at a redshift $z>0.007$ to look for spatial clustering.
Indeed, we identify a sub-group of 12 galaxies (including NGC~4602)\footnote{There are far more galaxies in the W-M sheet than this sub-sample of galaxies, but we were interested only in verifying that NGC~4602 is indeed associated with the W-M sheet, for which only a local subset of galaxies is needed.}
that appear clustered at similar redshifts, and are also within the 3D structural extent of the W-M sheet provided by \citet{kim+2016_virgocluster}.
We note that \citet{castignani+2022} also independently confirmed the W-M sheet structure, but their sample selection only extends down to declination $-$1.3$^\circ$, and thus does not include NGC~4602 which is at a declination of $-$5.13$^\circ$.

Of particular interest is the contribution to the DM of FRB~20200723B from this W-M sheet.
With an estimate of the distance that this burst traveled \textit{through} the W-M sheet to its front face (relative to Earth), we can thus use DM budgeting to place constraints on the free electron density of this filamentary structure.
We use the Cartesian supergalactic coordinates provided by \citet{kim+2016_virgocluster} to estimate the physical extent of the W-M sheet.
The supergalactic coordinate system is referenced to the ``supergalactic plane'', i.e., the plane of the Local Supercluster \citep{peebles_2022};
this is a great circle traced across the sky in the local Universe within which there appears to be a concentration of extragalactic matter \citep{deVaucouleurs1971}.
In the Cartesian supergalactic coordinate system, SGX and SGY axes are in the supergalactic plane, with the SGZ axis pointing towards the north supergalactic pole and the Earth at the origin.
The SGY axis, around the Virgo Cluster, approximately corresponds to the line of sight \citep{castignani+2022}.
These coordinates are often specified in units of $h^{-1}$~Mpc.

\begin{figure}[htbp]
    \centering
    \includegraphics[trim={0 1.5cm 0 0},clip, width=1.0\columnwidth]{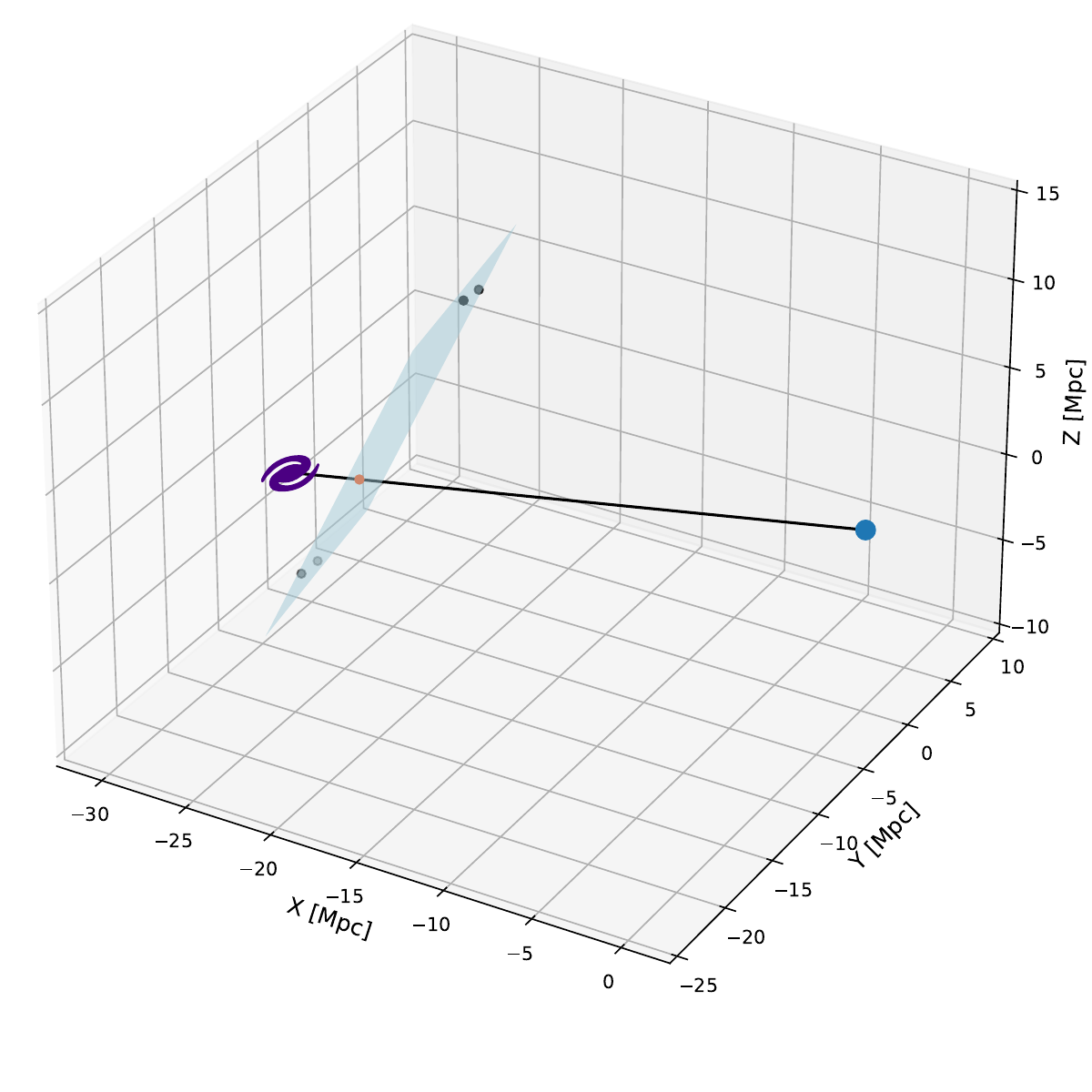}
    \caption{
        A 3D visualization, in Cartesian coordinates, of the best-fit plane (light blue sheet) of the front face of the W-M sheet between NGC~4602 (purple spiral galaxy icon) and Earth (blue dot).
        We emphasize that NGC~4602 is \textit{within} the W-M sheet, 
        and that the full volumetric extent of the W-M sheet, including its back face, is not depicted in this plot. 
        We use the redshift-independent Tully-Fisher distance measurement for NGC~4602.
        The plane is fit to the four points spanning the corners of front face of the W-M sheet (black dots);
        for added visualization effect, the plane is plotted slightly past these points.
        Along the line of sight between NGC~4602 and Earth (straight black line), the point of intersection with the best-fit plane of the front face of the W-M sheet is plotted in orange.
        The distance from NGC~4602 to the point of intersection is $3.9^{+2.9}_{-2.6}$~Mpc.
    }
    \label{fig:wmsheet_viz}
\end{figure}

The provided coordinate ranges of the W-M sheet are given as (SGX, SGY, SGZ) = ($-13.38 \sim -1.66$, $16.03 \sim 24.99$, $-3.10 \sim -1.10$) $h^{-1}$~Mpc.\footnote{This filamentary structure is elongated along the SGY axis, hence the designation ``W-M sheet'' instead of ``W-M filament'' \citep{kim+2016_virgocluster}.}
Towards the direction of the W-M sheet from Earth, the range of SGX--SGZ coordinates at the maximum SGY value roughly span a plane marking the start of the filamentary W-M sheet structure.
Adopting \textit{Planck} cosmological parameters \citep{planck2018vi} for the supergalactic coordinates, where $H_\mathrm{0} = 100 \ h$~km~s$^{-1}$~Mpc$^{-1}$ and $h =  0.6766$, we use singular value decomposition to obtain a best-fit plane to the coordinates of the W-M sheet closest to Earth.
From NGC~4602 along the line of sight, we can find the point of intersection with the front face of the W-M sheet (relative to Earth).
The distance FRB~20200723B would have traversed, from NGC~4602 through the W-M sheet, is thus $3.9^{+2.9}_{-2.6}$~Mpc.
A visualization of the geometry between NGC~4602, the front face of the W-M sheet relative to Earth, and the line of sight between NGC~4602 and Earth can be seen in Figure~\ref{fig:wmsheet_viz}.

\vspace{1em}
\subsection{DM budgeting}
\label{subsec:dm_budgeting}

We split the DM of FRB~20200723B into its line of sight components, aiming to place constraints on the DM contribution from the W-M sheet.
Given that NGC~4602 is in the local Universe, we ignore factors of $1+z$.
The DM for FRB~20200723B can thus be broken into
\begin{multline}
\label{eqn:dm_budget}
    \mathrm{DM} =
    \mathrm{DM}_\mathrm{mw,disk} + \mathrm{DM}_\mathrm{mw,halo}
    + \mathrm{DM}_\mathrm{igm} \\
    + \mathrm{DM}_\mathrm{int,halos}
    + \mathrm{DM}_\mathrm{sheet}
    + \mathrm{DM}_\mathrm{h}
\end{multline}
where DM$_\mathrm{mw,disk}$ is the interstellar medium (ISM) contribution from the Milky Way (MW);
DM$_\mathrm{mw,halo}$ is the MW halo contribution;
DM$_\mathrm{igm}$ is the diffuse intergalactic medium (IGM) contribution;
DM$_\mathrm{int,halos}$ is the contribution from intervening galaxy halos along the line of sight;
DM$_\mathrm{sheet}$ is the contribution from the W-M sheet;
and DM$_\mathrm{h}$ is the host galaxy contribution.
To maximize the simplicity of the methodology and the conservatism of the constraints, we adopt uniform  distributions (with very generous edges) for the probability distributions for each of these DM components.

For estimating the MW disk contribution to the DM of FRB~20200723B, there are two Galactic electron density models to consider --- the NE2001 \citep{ne2001} model and the YMW16 model \citep{ymw2016}.
Towards FRB~20200723B, the NE2001 prediction is $\sim$$33 \pm 7$~pc~cm$^{-3}$, while the YMW16 prediction is $\sim$$25 \pm 5$~pc~cm$^{-3}$.
We take the 2$\sigma$ value below the YMW16 prediction and the 2$\sigma$ value above the NE2001 prediction to denote the edges of our probability distribution for the MW disk contribution to the total DM.
Thus, DM$_\mathrm{mw,disk} \in [15, 47]$~pc~cm$^{-3}$.

Estimates for the MW halo contribution to the DM span $\approx$10--80~pc~cm$^{-3}$ \citep[e.g.,][]{dolag2015,prochaskazheng2019,keatingpen2020}.
As some halo models may overestimate MW halo DM contributions, as discussed by \citet{cook+2023_halo}, we conservatively adopt a range DM$_\mathrm{halo} \in [10, 80]$~pc~cm$^{-3}$.

As NGC~4602 is only $\approx$32~Mpc away, the diffuse IGM contribution to the DM is likely minimal.
Nevertheless, we follow the prescription of \citet{simha+2020} to estimate\footnote{The code used to estimate DM$_\mathrm{igm}$ can be found in the \texttt{FRBs} repository at \url{https://github.com/FRBs/FRB/blob/1eb95e0/frb/dm/igm.py\#L341}.} this contribution at $z=0.0085$ and find it is $\approx$4~pc~cm$^{-3}$.
We conservatively double this value to have a range DM$_\mathrm{igm} \in [0, 8]$~pc~cm$^{-3}$.

Gas from intervening galaxy halos along the line of sight can contribute to the DM of an observed FRB \citep[e.g.,][]{connor+2023, lee+2023_frb190520b_foreground, khrykin+2024_flimflam_dr1}.
Following the methodology used by \citet{simha+2020,simha+2021,simha+2023} and \citet{lee+2023_frb190520b_foreground}, we assume the gas content of galaxy halos extend out to twice their virial radii\footnote{Free electrons extending to twice the virial radii of galaxies is informed by recent hydrodynamical simulation results \citep[e.g.,][]{ayromlou+2023}.
For the purposes of our DM budgeting, this assumption also leads to a higher DM$_\mathrm{int,halos}$ upper limit than if we assume gas halos are truncated at the virial radius.}
and query the NED-LVS for galaxies with impact parameters $b_\perp<1$~Mpc and stellar mass estimates.
We then convert the stellar masses for these galaxies into halo masses using a stellar to halo mass ratio \citep{moster+2013},
adopt a modified Navarro-Frenk-White profile \citep[NFW;][]{nfw_profile} for the halo baryonic gas density,
and estimate the resulting DM contribution.
The resulting average intervening halo contribution for FRB~20200723B is $\approx$13~pc~cm$^{-3}$.
As above, we conservatively double this value to have a range DM$_\mathrm{int,halos} \in [0, 26]$~pc~cm$^{-3}$.

Taking the minimum contribution of all the above components, we are left with a maximum possible contribution of $\approx$219~pc~cm$^{-3}$ for the combined contribution of DM$_\mathrm{sheet}$ and DM$_\mathrm{h}$.
Using the prescription of \citet{simha+2020,simha+2021,simha+2023}, we find that the halo alone of NGC~4602 could contribute up to $\sim$80~pc~cm$^{-3}$.
Given this modeling, NGC~4602 is likely to contribute DM$_\mathrm{h} > 80$~pc~cm$^{-3}$ from its total combined halo and ISM contributions.
Motivated by these numbers, we assume a minimal host galaxy DM contribution from NGC~4602 of 40~pc~cm$^{-3}$, half the amount from its expected halo contribution alone.\footnote{Assuming a minimum DM$_\mathrm{h} = 0$~pc~cm$^{-3}$ is less physically motivated and does not qualitatively affect any conclusions.}
In other words, we obtain DM$_\mathrm{h} \in [40, 219]$~pc~cm$^{-3}$.
Thus, we arrive at a DM contribution range from the W-M sheet of DM$_\mathrm{sheet} \in [0, 179]$~pc~cm$^{-3}$.

\subsection{Column density of the W-M sheet}
\label{subsec:wm_sheet_coldensity}

After the conservative DM budgeting in Section~\ref{subsec:dm_budgeting}, we have a plausible range of DM contribution values from the W-M sheet to the total observed DM of FRB~20200723B.
The upper limit DM contribution from the filamentary structure is DM$_\mathrm{sheet}~<~179$~pc~cm$^{-3}$, which is equivalent to a column density $\Sigma~<~5.5~\times~10^{20}$~cm$^{-2}$.
We can compare this column density to gas column densities of filamentary structures found in simulations.
The IllustrisTNG project consists of a suite of cosmological magneto-hydrodynamical simulations \citep{pillepich+2018_illustrisTNG}.
The three primary simulations are TNG50, TNG100, and TNG300, where the numbers correspond to side length, in Mpc, of the cubic volumes in the simulation box sizes.
Figure~2 of \citet{marinacci+2018_illustrisTNG_firstresults} show thick slice snapshots from the TNG100 and TNG300 runs.
We see a range of column densities $\Sigma = 5~\times~10^{19}$~cm$^{-2}$ to $5~\times~10^{20}$~cm$^{-2}$ are expected for cosmic filamentary structures,
consistent with the column density we find through the W-M sheet.

We can also compare this column density with the stacked Planck $y$ map that \citet{tanimura+2020} made and used to detect the thermal Sunyaev-Zel'dovich (tSZ) signal in filaments.
The Compton $y$ parameter quantifies the amplitude of the tSZ signal via the electron pressure integrated along the line of sight,
\begin{equation}
    y = \frac{\sigma_T}{m_e c^2} \int k_\mathrm{B} n_e T_e dl.
\end{equation}
Here $\sigma_T$ is the Thomson cross section, $m_e$ is the electron mass, $c$ is the speed of light, $k_\mathrm{B}$ is the Boltzmann constant, $n_e$ is the free electron number density, and $T_e$ is the temperature.
From their Figure~5, we take a Compton $y$ parameter at the radial center of a filament as $y \approx 2 \times 10^{-8}$.
If we take our upper limit column density $\int n_e dl < 5.5 \times 10^{20}$~cm$^{-2}$ and combine it with this Compton $y$ parameter, then we can estimate a lower limit on the temperature in the W-M sheet to be $T_e > 3 \times 10^5$~K.
This limit is consistent with the cosmic web filament gas temperature measured by \citet{tanimura+2020} ($\approx$10${^6}$~K).

\subsection{Free electron density of the W-M sheet}
\label{subsec:wm_sheet_n_e}

We also consider the free electron number density.
Using the upper limit filamentary DM contribution of 179~pc~cm$^{-3}$, and taking the distance traversed from NGC~4602 through the W-M sheet as $3.9^{+2.9}_{-2.6}$~Mpc (Section~\ref{subsec:wm_sheet}), we can set an upper limit on the mean free electron density through the W-M sheet: $\langle n_e \rangle~<~4.6^{+9.6}_{-2.0} \times 10^{-5}$~cm$^{-3}$.
It is of course possible that the medium in the W-M sheet is notably anisotropic, with regions of significant overdensities and underdensities.
It is thus also possible that the medium composition in the W-M sheet in front of NGC~4602 (relative to us) significantly differs from the W-M sheet medium behind NGC~4602.
Depending on the exact makeup of the W-M sheet, the true free electron constraint could be different.
However, the uncertainty in the Tully-Fisher distance to NGC~4602 is large ($\sim$9\% relative error), which means the uncertainty in the average sightline free electron density is large.
Therefore, we assume that this large uncertainty encompasses any possible medium fluctuations that would affect the free electron density constraint.

To assess the consistency of $\Lambda$CDM with our free electron number density limit through the W-M sheet, we can compare that result to $n_{e,0}$, the free electron density of the Universe at $z=0$.
In a fully ionized universe, $n_{e,0}$ is defined as
\begin{equation}
    n_{e,0} = f_{d,0} \left( \mathrm{Y}_\mathrm{H} + \frac{1}{2} \mathrm{Y}_\mathrm{He} \right)
    \frac{3 H_0^2 \Omega_{b,0}}{8 \pi G m_p}
\end{equation}
where $f_{d,0}$ is the fraction of baryons in diffuse ionized gas at $z=0$,
$\Omega_{b,0}$ is the $\Lambda$CDM baryon density parameter at $z=0$,
$G$ is the gravitational constant,
$m_p$ is the mass of the proton,
$H_0$ is the Hubble constant,
and Y$_\mathrm{H}$ and Y$_\mathrm{He}$ are the primordial hydrogen and helium abundances, respectively.
Using \citet{planck2018vi} values, $\mathrm{Y}_\mathrm{H} + \mathrm{Y}_\mathrm{He}/2 \approx 0.875$ and $\Omega_{b,0} = 0.04897$.
For the fraction of baryons in diffuse ionized gas, we adopt $f_{d,0} = 0.84$ \citep[e.g.,][]{simha+2020, james+2022_h0}.
With these values, we obtain a free electron density of $n_{e,0}~\approx~2~\times~10^{-7}$~cm$^{-3}$ for the $z=0$ Universe.\footnote{Using multiple FRB sightlines, \citet{khrykin+2024_flimflam_dr1} measure the fraction of baryons in the IGM --- excluding gas within halos --- to be $f_{d,0} = 0.63^{+0.09}_{-0.07}$. This value leads to $n_{e,0}~=~4.0^{+0.57}_{-0.44}~\times~10^{-8}$~cm$^{-3}$.}
The upper limit of our free electron density through the W-M sheet is thus no more than $\approx$2 orders of magnitude more overdense than the mean free electron density of the Universe.
As one would expect a baryonic structure to be an overdensity, this result can be interpreted as our free electron density upper limit through the W-M sheet being consistent with $\Lambda$CDM expectations.

Our free electron density result from FRB~20200723B through the W-M sheet can also be compared to IllustrisTNG simulation results.
\citet{martizzi+2019} use snapshots from the TNG100 simulations to study baryons throughout various cosmic web structures at $0 \leq z \leq 8$.
They show, in their Figure~4, a phase diagram of baryons at $z=0$ in filaments and sheets.
We see in their results that, for filaments and sheets at $z=0$, the highest probability density of baryons in the warm-hot intergalactic medium (WHIM, $10^5$~K$ < T < 10^7$~K), which would dominate in filamentary structures, is around a baryonic number density of $n_\mathrm{H} \sim 10^{-6}$~cm$^{-3}$ (baryonic number density can be taken as a rough proxy for electron number density, as on large scales, electrons trace baryons).
\citet{galarraga-espinosa+2021} use the TNG300-1 simulation box at $z=0$ to study gas phases around cosmic filamentary structures.
They also find that for the WHIM in filamentary structures,
the highest probability density of baryons in the WHIM is just over $n_\mathrm{H} \sim 10^{-6}$~cm$^{-3}$ (e.g., their Figures 4 and 9).
These baryonic number densities of $n_\mathrm{H} \sim 10^{-6}$~cm$^{-3}$ are consistent with our upper limit $\langle n_e \rangle~<~4.6^{+9.6}_{-2.0} \times 10^{-5}$~cm$^{-3}$.
This average sightline free electron density constraint can also be compared to the analysis done by \citet{erciyes+2023}.
They use a non-detection of \textit{Planck} tSZ cross-correlations with local Universe galaxies to estimate the average electron pressure; their upper limit on a volume-averaged filament electron number density is higher at $\langle n_e \rangle \lesssim 4 \times 10^{-4}$~cm$^{-3}$.

\section{Scattering timescale considerations}
\label{sec:scattering}

The scattering timescale $\tau$ of a radio wave, also referred to as the scatter-broadening or pulse broadening time, results from multi-path propagation through inhomogeneities in intervening media.
Understanding whether scattering is dominated by the local environment to the FRB \citep[e.g.,][]{ocker+2023_scatteringvariability} can give insight into progenitor models.
Scattering measurements or limits combined with sightlines through media such as intervening galaxy halos can provide insights into the gas properties of these environments \citep[e.g.,][]{prochaska+2019_halo, connor+2020, ocker+2021_halos}.

For scattering, a frequency-dependent pulse broadening effect can be observed where the resulting one-sided exponential scattering tail follows $\tau \propto \nu^{\alpha}$.
As mentioned in Section~\ref{subsec:burst_props},
we adopt $\alpha=-4$, the spectrum predicted by refractive scattering 
\citep[e.g.,][]{rickett1977}.

For this section, we adopt a model where the scattering medium is composed of ionized cloudlets embedded within more diffuse gas, a model commonly adopted for the CGM \citep[e.g.,][]{tumlinson+2017}.
(In Section~\ref{subsec:scattering_wm_sheet}, we discuss another model interpretation as well.)
In such a cloudlet model,
the degree of fluctuations can be parameterized by the density fluctuation parameter \citep{cordes+2016}
\begin{equation}
\label{eqn:F}
    \widetilde{F} =
    \frac{\zeta \epsilon^2}
    {f_\mathrm{v}(l_{\rm o}^2 l_{\rm i})^{1/3}} \ \ (\mathrm{pc}^2 \ \mathrm{km})^{-1/3}
\end{equation}
where 
$\zeta$ is the mean density variation between cloudlets,
$\epsilon$ is the fractional rms of density fluctuations within a cloudlet,
$f_\mathrm{v}$ is the volume filling factor,
and $l_{\rm o}^2$ (pc) and $l_{\rm i}$ (km) define the outer and inner scales of the density fluctuations.
A formalism developed by \citet{cordes+2016}, \citet{ocker+2021_halos}, and \citet{cordes+2022_dmtau} relates these electron density fluctuations to scattering timescales observed for pulsars and FRBs.
This formalism also allows for the geometry of intervening scattering layers boosting the observed scattering timescales via a geometric leverage factor $G_\mathrm{scatt}$.
For extragalactic distances where the FRB source is at redshift $z_\mathrm{s}$ and the intervening scattering layer is at redshift $z_\ell < z_\mathrm{s}$, we can define
\begin{equation}
\label{eqn:G}
    G_\mathrm{scatt}(z_\mathrm{s}, z_\ell) = \frac
    { 2 d_\mathrm{sl} d_\mathrm{lo} }
    { L d_\mathrm{so} }.
\end{equation}
Here, $d_\mathrm{sl}$ is the angular diameter distance from the source to the layer, $ d_\mathrm{lo}$ is the distance from the layer to the observer, and $d_\mathrm{so}$ is the distance from the source to the observer.
For galaxy disks, the scattering layer thickness can be taken as $L = 1$~kpc \citep{cordes+2022_dmtau}.
When the scattering layer is in the FRB host galaxy as opposed to an intervening galaxy, the geometric leverage factor is often simplified to $G_\mathrm{scatt}=1$.
This simplification comes from $d_\mathrm{sl} \rightarrow L/2$ and is thus agnostic as to whether the scattering screen is local ($\sim$pc) to the FRB as opposed to just within the host galaxy ($\sim$kpc), as long as the scattering layer thickness is twice the distance from the source to the center of the scattering layer.
In practice, however, $G_\mathrm{scatt} = 1$ can be a poor approximation ---
e.g., for pulsars, which reside in the same galaxy as their scattering media, their scattering is dominated by thin, highly localized screens where $L \ll d_\mathrm{sl}$ \citep{brisken+2010}.
While more complicated geometric configurations are not considered in this work, we acknowledge that they are a physical possibility.

In this model, a burst that travels through a layer contributing $\mathrm{DM}_\ell$ (in the rest frame of the scattering medium) with a geometric leverage factor $G_\mathrm{scatt}$ has an observed scattering time of
\begin{multline}
\label{eqn:tau_afg}
    \tau(\mathrm{DM}_\ell,\nu, z_\mathrm{s}, z_\ell) \simeq 
    (48 \ {\rm ns}) \ \times \\
    \ A_\tau \ \nu^{-4} (1+z_\ell)^{-3} \ \widetilde{F} \ G_\mathrm{scatt}(z_\mathrm{s}, z_\ell) \ \mathrm{DM}_\ell^2
\end{multline}
where $\nu$ is the observing frequency in GHz \citep{ocker+2022_horizons}.
The parameter $A_\tau$ parameterizes the shape of the pulse-broadening function; henceforth we adopt $A_\tau=1$. 
The derivation of the pre-factor can be found in the Appendix of \citet{cordes+2022_dmtau}.
We note that this ionized cloudlet model assumes pulse scatter-broadening is primarily due to diffractive scattering of a volume-filling Kolmogorov spectrum of density fluctuations.
However, it is possible that scattering is dominated by refractive scattering due to a small number of intermittent structures with large density variations \citep{jow+2024_cgmsheet}.
This model assumption should be kept in mind when interpreting subsequent inferences made in this section.

Assuming both FRB~20200723B and its primary scattering screen originate from within NGC~4602 at $z_\ell = 0.0085$, we take the geometric factor to be $G_\mathrm{scatt}=1$.
Using Equation~\ref{eqn:tau_afg} and our conservative range of host galaxy DMs (Section~\ref{subsec:dm_budgeting}), we can thus constrain the degree of fluctuation in the scattering medium $\widetilde{F}$.
We assume the DM contributed by the scattering medium layer is equivalent to the host DM \citep{cordes+2022_dmtau}.
The range of DM$_\mathrm{h} \in [40, 219]$~pc~cm$^{-3}$ therefore corresponds to a range $\widetilde{F} \in [390, 13]$~(pc$^2$ km)$^{-1/3}$, respectively.

It appears that $\widetilde{F} \lesssim 1$~(pc$^2$ km)$^{-1/3}$ for most FRBs for which $\widetilde{F}$ is estimated \citep[e.g.,][]{cordes+2022_dmtau}.
Additionally, based on Milky Way pulsar observations, the maximum expected $\widetilde{F}$ for the thin disk of a spiral galaxy is about 1~(pc$^2$ km)$^{-1/3}$ \citep{ocker+2022_horizons}.
However, higher $\widetilde{F}$ values are possible within the thick disk \citep{ocker+2021_halos}.
While FRB~20200723B appears to have an unusually high fluctuation parameter, it is not an infeasible amount when considering Milky Way observations of pulsars.
Furthermore, qualitatively, it is possible that there is a high degree of density fluctuations associated with the large observed scattering timescale of FRB~20200723B.

\subsection{Can the scattering be explained by the W-M sheet?}
\label{subsec:scattering_wm_sheet}

Cosmic filamentary structures between galaxy clusters are thought to be composed mostly of $T \gtrsim 10^6$~K gas in the WHIM \citep[e.g.,][]{deGraaff+2019, tanimura+2020, galarraga-espinosa+2021}.
Compared to cooler and denser gas such as, e.g., gas cloudlets in the CGM of galaxies at $T \sim 10^4$~K, the WHIM is not thought to significantly contribute to the temporal scattering of radio waves \citep[e.g.,][and references therein]{prochaski+2013, hennawi+2015, tumlinson+2017, mccourt+2018, VedanthamPhinney2019}.
We can verify the plausibility of whether FRB~20200723B can have its large scattering timescale fully explained by passing through an overdensity at the front of the W-M sheet (relative to Earth).

Taking the redshift-independent distance of NGC~4602 and its distance to the front of the W-M sheet,
we can derive a geometric factor $G_\mathrm{scatt} \approx 6.8 \times 10^3$ (Equation~\ref{eqn:G}).
This is the largest possible geometric leverage factor, as it assumes a configuration where all the scattering is attributable to an overdense layer at the furthest possible distance from NGC~4602 while still being in the W-M sheet, thus maximizing the distance from the source to the scattering layer.
Considering Equation~\ref{eqn:F}, we note that the volume filling factor of the WHIM can be up to $f_\mathrm{v} \sim 0.5$ \citep{nevalainen+2015} due to its temperature, compared to typical values of $f_\mathrm{v} \sim 10^{-3} - 10^{-2}$ for cooler $10^4$~K gas commonly invoked to explain scattering \citep{VedanthamPhinney2019, ocker+2021_halos}.
Using, e.g., a two orders-of-magnitude higher value of $f_\mathrm{v}$ to reflect the WHIM in Equation~\ref{eqn:F}, we obtain a two orders-of-magnitude lower $\widetilde{F}$ parameter.
If we take $\widetilde{F} = 10^{-2}$~(pc$^2$ km)$^{-1/3}$ \citep[two orders of magnitude lower than upper range of typical values of $\widetilde{F}$ for FRBs;][]{cordes+2022_dmtau},
the largest possible geometric leverage factor boosting the radio wave scattering in this scenario,
and all the possible DM$_\mathrm{sheet}$ contributing to the scattering, we can obtain a scattering timescale of $\tau_\mathrm{400\,MHz} \sim 4$~s from the W-M sheet.
While this timescale can be sufficient to explain the observed scattering timescale, any factor of 4 suppression in the relevant parameters --- e.g., a lower $\widetilde{F}$ parameter, a smaller geometric leverage factor, or a smaller proportion of DM$_\mathrm{sheet}$ contributing to the scattering, all of which are very possible --- means gas from the W-M sheet alone cannot account for the full scattering timescale of FRB~20200723B.

So far we have assumed the cloudlet model discussed at the start of Section~\ref{sec:scattering}.
However, it is possible that scattering structures within the CGM are better described by a sheet-like geometry, as proposed by \citet{PenLevin2014_sheets} and \citet{jow+2024_cgmsheet} based on observations of pulsar and FRB scintillation.
Sheet-like geometries are often invoked to explain why radio sources could exhibit scatter-broadening due to multipath propagation without a large number of images suppressing scintillation \citep{jow+2024_cgmsheet}.
Edge-on sheets aligned with the line of sight of the FRB would lead to a huge boost in spatial density fluctuations without suppressing FRB scintillation, while also remaining consistent with observations from quasar absorption studies of CGM \citep{jow+2024_cgmsheet}.
We can consider whether such a sheet-like geometry within the W-M sheet filamentary structure can explain the observed scattering timescale of FRB~20200723B.
However, we note that the temperature limit $T_e > 3 \times 10^5$~K we obtained in Section~\ref{subsec:wm_sheet_coldensity} for the W-M sheet already suggests that the sheet is unlikely to dominate the scattering, as the high temperature makes it unlikely for matter to clump enough to give the amount of density fluctuations needed for large scatter-broadening timescales.

To assess the plausibility of the scenario where the scattering is dominated by a refractive plasma lens in the W-M sheet, we note that the refractive scattering timescale can be written as \citep{jow+2024_cgmsheet}
\begin{multline}
    \label{eqn:tau_sheet}
    \tau_\mathrm{400\,MHz} = 0.14 \, \mu\mathrm{s} \, (1+z_\ell)^{-3} \\ 
    \times 
    \left( \frac{d_\mathrm{sl} d_\mathrm{lo}/d_\mathrm{so}}{\mathrm{Mpc}} \right)
    \left( \frac{N_e r}{\mathrm{cm}^{-3}} \right)^{2}
    \left( \frac{f}{\mathrm{MHz}} \right)^{-4}
\end{multline}
where $r$ is the size of the refractive structure transverse to the line of sight, and for a spherical geometry, $N_e r$ is the effective electron density of the lens.
Placing the lens at the furthest edge of the W-M sheet, in order to obtain the observed scattering time, one would require an excess electron density of $n_e \sim 10^3$~cm$^{-3}$.
Given our upper-bound on the electron density of the W-M sheet ($\langle n_e \rangle~<~4.6^{+9.6}_{-2.0} \times 10^{-5}$~cm$^{-3}$), this corresponds to an overdensity a factor of $10^8$ larger than the ambient medium.
While an inclined sheet-like or filamentary geometry may enhance the effective projected column density, this enhancement is only as large as the aspect ratio, $A$, of the filament or sheet (the quantity $N_e r$ is at most $n_e A$).
Thus, in order to reproduce the observed scattering timescale, the refractive sheet or filament would need to be stretched by at least a factor of $10^8$ relative to its thickness --- a rather unlikely scenario.

Nonetheless, it would be interesting for future work to consider other models of scattering from media within filamentary structures, particularly one that could explain the large scattering timescale of FRB~20200723B.

\subsection{Can the scattering be explained by an unseen true host galaxy lying behind NGC~4602?}
\label{subsec:scattering_background}

Given the combination of high scattering timescale of FRB~20200723B, as well as the much closer distance of NGC~4602 than is naively implied by the total observed extragalactic DM (Section~\ref{sec:localization}), one may ask whether it is plausible that the FRB originates from a host galaxy \textit{behind} NGC~4602.
Setting aside chance alignment probability arguments against the likelihood of such a scenario, we can consider an optimistic scenario where the disk of NGC~4602 is the scattering layer for FRB~20200723B.

Referring to Equations~\ref{eqn:G} and \ref{eqn:tau_afg},
we see that the combination of fluctuation parameter $\widetilde{F}$, geometric leverage factor $G_\mathrm{scatt}$, and DM of the scattering layer DM$_\ell$ must be consistent with a given scattering timescale.
In order to minimize the impact of the geometric leverage factor, we consider an unseen background galaxy only 10~Mpc behind NGC~4602;
much further distances are allowed out to $z_\mathrm{max}$ (e.g., Section~\ref{sec:localization}), but would generally only increase the geometric factor (Equation~\ref{eqn:G}).\footnote{We argue that given the low DM of FRB~20200723B, a background galaxy with a sightline intersecting the halo of NGC~4602 would likely be visible in archival Legacy Survey imaging. Indeed, in Figure~\ref{fig:bb_overlaid}, the visible background galaxies have photometric redshifts that place them $\gg$10~Mpc behind NGC~4602.}
For similar reasons, we also consider a rather low DM$_\ell$ contribution of  40~pc~cm$^{-3}$ from NGC~4602 (e.g., Section~\ref{subsec:dm_budgeting}).
In this scenario, such a configuration leads to a geometric leverage factor $G_\mathrm{scatt} \approx 1.5 \times 10^4$ and a fluctuation parameter $\widetilde{F} \simeq 0.026$ --- a not physically implausible value when compared to Milky Way observations of pulsars through the thick disk \citep[e.g.,][]{cordes+2022_dmtau,ocker+2022_horizons,faber+2024_scattered}.
Thus, we have a possible geometric configuration that can reproduce the observed scattering timescale.

However, a putative value $\widetilde{F} = 1$ is more likely for the thin disk of a spiral galaxy, and one would expect a much larger DM$_\ell$ contribution from a sightline through the ISM of a spiral galaxy as massive as NGC~4602.
To compare with the Milky Way, the NE2001 model applied at the same Galactic longitude as FRB~20200723B but at a Galactic latitude $b = 15^\circ$ gives a DM of $\sim$135~pc~cm$^{-3}$.
As a point of comparison, if we set $\widetilde{F} = 1$ and keep the background galaxy at just 10~Mpc behind NGC~4602, we find that a DM$_\ell$ contribution of $\approx$3.4~pc~cm$^{-3}$ is necessary to obtain the observed scattering timescale, far lower than an expected DM contribution from a sightline through a Milky Way-like Galactic disk, and even through Milky Way-like halos.
It is also possible for a background galaxy to lie $\gg$10~Mpc behind NGC~4602;
increasing either $G_\mathrm{scatt}$ or DM$_\ell$ any further would lead to a scattering timescale significantly larger than the actual observed scattering timescale of FRB~20200723B.

Thus while in this formalism, the DM and $\tau$ measurements of FRB~20200723B alone cannot rule out the possibility of the burst originating from an unseen background galaxy and being scattered through the disk of NGC~4602, such a scenario requires a contrived combination of $\widetilde{F}$, DM$_\ell$, and chance alignment of a background galaxy.

\subsection{Scattering from within NGC~4602}
\label{subsec:scattering_ngc4602}

The physically simplest explanation for the scattering timescale of FRB~20200723B is via a scattering layer from within NGC~4602.
\citet{bhardwaj+2024_inclination} note that there is empirical evidence for a selection bias against FRBs from host galaxies with high inclination angles, and propose large scattering timescales beyond detectability as the cause for this bias, although this correlation remains to be tested with scattering measurements and future studies of telescope systematics.
FRB~20200723B, with its high scattering timescale and putative host galaxy NGC~4602, with a high inclination angle $i = 73 \pm 3^\circ$ \citep{kourkchi+2020_cosmicflows}, is consistent with the paradigm proposed by \citet{bhardwaj+2024_inclination}.
However, for FRB~20200723B, we caution against over-interpretation from host galaxy inclination angle alone --- while highly scattered FRBs may preferentially originate from galaxies with high inclination angles, not every highly scattered FRB \textit{must} originate from one.
Some local FRB environments are known to be associated with extreme, dynamic local environments \citep[e.g.,][]{michilli+2018_121102, ocker+2023_scatteringvariability};
thus, local environments, irrespective of galaxy orientation, could contribute significantly to the scattering timescale.
Additionally, not every highly inclined galaxy will emit only highly scattered FRBs.
Indeed, FRB~20210603A, a burst localized at detection using VLBI from untargeted observations, is in a galaxy with inclination angle $i = 83 \pm 3^\circ$ and has a significantly smaller scattering timescale at $\tau_\mathrm{600\,MHz} \lesssim 165 \pm 3~\mu$s \citep{CLS+2023_first_realtime_frb_vlbi}.
\citet{CLS+2023_first_realtime_frb_vlbi} note that their observations and inferred parameters are consistent with FRB~20210603A originating from within its host galaxy disk.

\begin{figure}[htbp]
    \centering
    \includegraphics[width=1.0\columnwidth]{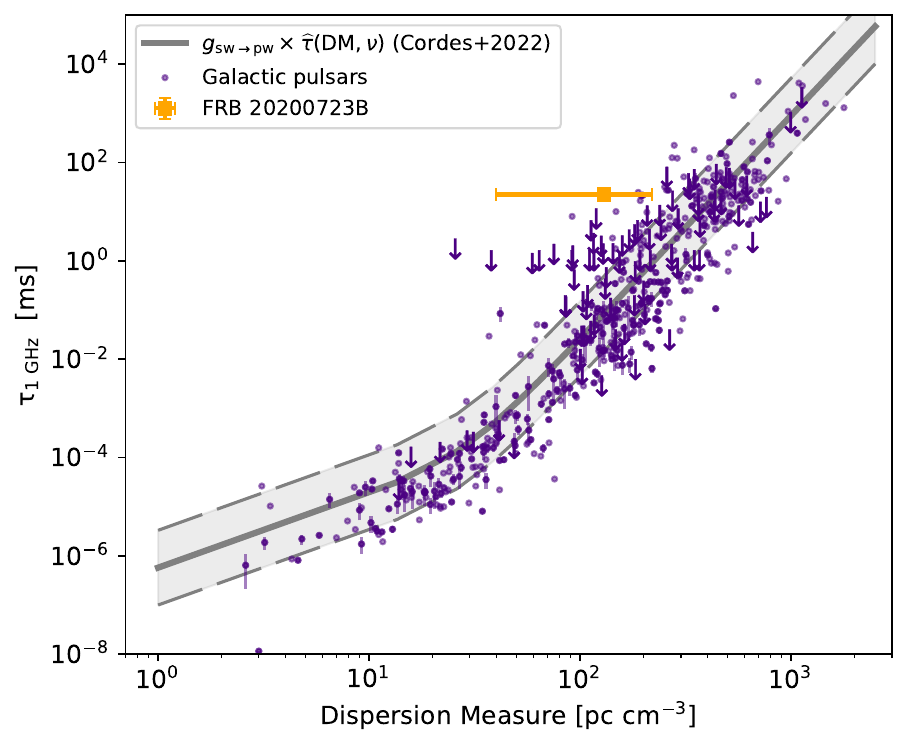}
    \caption{
        The empirical mean scattering--DM relation (Eqn.~\ref{eqn:taudm_empirical}), multiplied by a factor $g_\mathrm{sw \rightarrow pw} = 3$ to account for scattering for FRBs, is plotted in gray with the $\pm$1$\sigma$ region shaded around the line.
        Scattering and DM data for 570 Galactic pulsars (484 averages, 86 upper limits) are also plotted in indigo;
        these data are provided courtesy of J. Cordes, and are also published in \citet{cordes+2022_dmtau}.
        FRB~20200723B is plotted in orange, scaled to its nominal value at 1~GHz using the best-fit scattering timescale and index from Section~\ref{subsec:burst_props}; horizontal error bars denote the range of DM contributed by the host galaxy in its rest frame.
        At the maximum host DM, the observed scattering timescale for FRB~20200723B is $\sim$2$\sigma$ greater than what would be expected from this relation (i.e., scattering from a Milky Way-like galaxy).
        Decreasing the host DM increases the excess of scattering relative to the empirical relation.
    }
    \label{fig:hockey_stick}
\end{figure}

NGC~4602 is a galaxy with a large number of prominent H\,\textsc{ii} regions \citep{tsvetanov+1995}.
H\,\textsc{ii} regions are known to contribute to the observed scattering of bursts from pulsars \citep[e.g.,][]{mall+2022, ocker+2024_hii, ocker+2024_bowshocks}.
For pulsars in the Milky Way, there is an empirical mean scattering--DM relation:
\begin{multline}
\label{eqn:taudm_empirical}
    \widehat{\tau}(\mathrm{DM},\nu) = (1.9\times10^{-7} \rm~ms)~\nu^{-4}DM^{1.5} \\
    \times (1 + 3.55\times10^{-5}\rm DM^{3.0})
\end{multline}
where the frequency $\nu$ is 1~GHz and $\tau \propto \nu^{-4}$ was adopted as the fiducial scaling to obtain this relation \citep{cordes+2022_dmtau}.
The scatter for this mean relation is $\sigma( \log_{10} \tau ) = 0.76$.
For FRBs, this $\widehat{\tau}(\mathrm{DM},\nu)$ relation is multiplied by a factor $g_\mathrm{sw \rightarrow pw} = 3$ to account for how scattering from host galaxy interstellar media is larger for FRBs than for Galactic pulsars.
This factor is because wavefronts from extragalactic FRBs present as plane waves, while wavefronts from Galactic pulsars are spherical \citep{cordes+2022_dmtau}.
Using this relation and assuming the maximum host DM contribution of DM$_\mathrm{h} = 219$~pc~cm$^{-3}$, at 1~GHz, FRB~20200723B would still have significant ($\sim$2$\sigma$) excess scattering compared to its DM.
FRB~20200723B is plotted against the empirical mean--scattering DM relation for FRBs and measurements from Milky Way pulsars in Figure~\ref{fig:hockey_stick}.

However, \citet{ocker+2024_hii} found that most of $\sim$30 identified pulsars with significant excess scattering ($>$2$\sigma$),
with DMs ranging from tens to hundreds of units, are spatially coincident with ionized foreground structures such as H\,\textsc{ii} regions.\footnote{For example, \citet{ocker+2024_hii} present J1730--3350 with a total DM of 261~pc~cm$^{-3}$ and scattering timescale $\tau_\mathrm{1\,GHz} = 27.3$~ms.
These values are pleasingly comparable to those of FRB~20200723B, with a maximum host galaxy DM of 219~pc~cm$^{-3}$ and scattering timescale $\tau_\mathrm{1\,GHz} = 22.3$~ms (scaled assuming $\tau \propto \nu^{-4.3}$ from the \texttt{fitburst} fit).}
It is also worth noting that \citet{ocker+2024_hii} identified $>$100 pulsars that intersected H\,\textsc{ii} regions within our Milky Way;
thus, qualitatively, it appears to be not uncommon for pulsars to have sightlines that intersect H\,\textsc{ii} regions in inclined galaxies.
As such, it is entirely plausible that FRB~20200723B has its high scattering timescale as a result of intersecting an H\,\textsc{ii} region along its line of sight.
However, in the absence of a sub-arcsecond localization illuminating the true sightline path of FRB~20200723B, an association with a visible H\,\textsc{ii} region cannot be unambiguously made.

Additionally, while plausible that a foreground H\,\textsc{ii} region is the source of the large scattering timescale of FRB~20200723B, we note there are other Galactic pulsars that lie above the mean scattering--DM relation due to reasons other than foreground H\,\textsc{ii} regions.
Such sources include the Crab Pulsar \citep{staelin+1968, deneva+2024, krishnakumar+2015} and PSR J0540$-$6919 \citep{seward+1984}, which both lie within supernova remnants, and PSR B1957+20 \citep{fruchter+1988, wang+2023, kuzmin+2007}, which is an eclipsing binary pulsar.\footnote{Most measurements and references are obtained from the ATNF Pulsar Catalogue at \url{https://www.atnf.csiro.au/research/pulsar/psrcat/} \citep{manchester+2005_atnf}.}
These pulsars have more local sources of scattering (supernova remnant, binary companion) relative to foreground H\,\textsc{ii} regions.
Thus, a broad diversity of physical scenarios are consistent with the observed DM and scattering of FRB~20200723B.

Beyond H\,\textsc{ii} regions, supernova remnants, or eclipsing binary pulsars, the observed scattering timescale could also be explained by
an unusually tumultuous circumburst environment, though we note the low RM upper limit found in Section~\ref{subsubsec:burst_props_pol} indicates the circumburst environment is likely not extremely dense and/or strongly magnetized.
Future observations and localizations with CHIME/FRB Outriggers \citep{lanman+2024_kko} will be promising for discovering whether there exist FRBs with high scattering timescales and localizations that coincide with extragalactic H\,\textsc{ii} regions.

\subsection{A comparison with injections}
\label{subsec:compare_injections}

In the first CHIME/FRB catalog, \citet{catalog1} noted that CHIME/FRB has a strong observational selection bias against FRBs with a scattering timescale above 10~ms at 600~MHz.
This conclusion was reached using the sample of synthetic bursts from an injections campaign in 2020~August, which aimed to characterize which bursts would have been detected as a function of their burst parameters \citep{merryfield+2023_injections}.
With such a high scattering timescale, FRB~20200723B should be very difficult for CHIME/FRB to observe, which raises the questions:
Why might FRB~20200723B have been detected in the first place, and can we investigate whether FRB~20200723B-like sources are intrinsically rare?
We thus aim to use these injections data\footnote{The full 2020 August injections data sets, as well as a data usage tutorial, are available at \url{https://chime-frb-open-data.github.io/}.} to contextualize the detection of FRB~20200723B by CHIME/FRB.
In this section, we use the \texttt{fitburst} measurements obtained with fixing the scattering index to $\alpha = -4$ (Section~\ref{subsec:burst_props}) in order to more properly compare with other CHIME/FRB observations.

\begin{figure}[htbp]
    \centering
    \includegraphics[width=1.0\columnwidth]{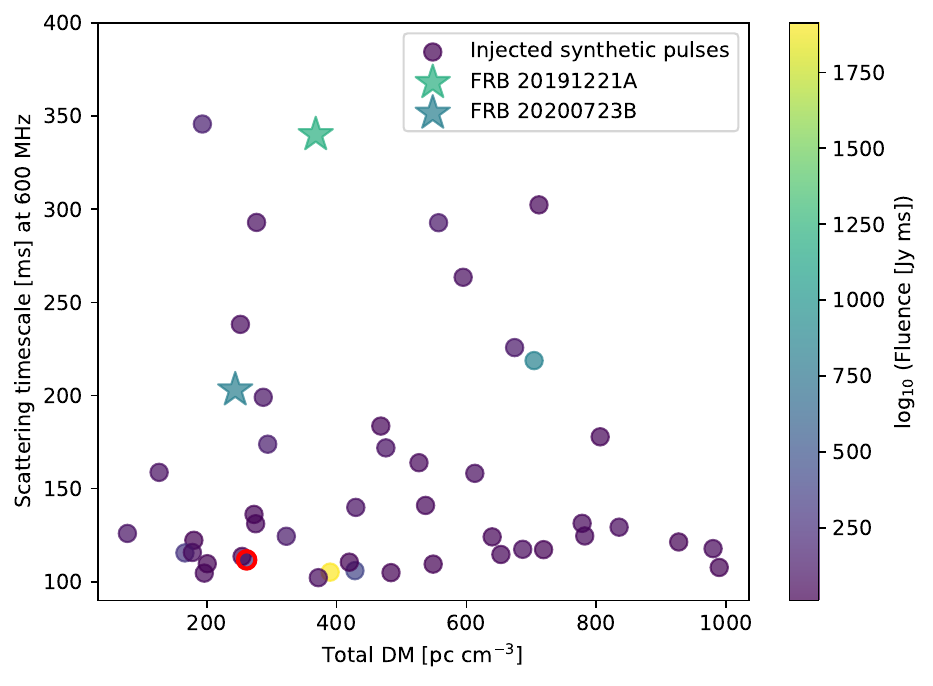}
    \caption{
        The total DM and scattering timescales of FRB~20191221A and FRB~20200723B, both marked with stars, plotted against the total DM and scattering timescales of bursts from the 2020~August CHIME/FRB injections campaign, plotted as circles.
        Only bursts injected with a scattering timescale $>$100~ms are plotted.
        The marker colors for the injected bursts logarithmically scale with their injected fluence values.
        For the two real observed FRBs, their marker colors correspond to their fluence lower limit values.
        The only injections burst detected by the CHIME/FRB detection pipeline is circled in red.
        There are no injected bursts for which the scattering timescale, fluence, and DM values are comparable to those of FRB~20191221A or FRB~20200723B, greatly limiting our ability to contextualize these observations against CHIME/FRB selection biases.
    }
    \label{fig:injections_comparison}
\end{figure}

Scaled to 600~MHz, the scattering timescale of FRB~20200723B is $\tau_\mathrm{600\,MHz} \sim$200~ms, and as noted in Section~\ref{subsec:burst_props}, it has a fluence lower limit $\gtrsim$820~Jy~ms.
We plot this event, as well as the highly scattered source exhibiting sub-second periodicity FRB~20191221A (with $\tau_\mathrm{600\,MHz} \sim$340~ms and fluence $\gtrsim$1200~Jy~ms), against bursts that were injected during the 2020 August injections campaign (Figure~\ref{fig:injections_comparison}).
We note that both of these bursts have scattering timescales and fluence lower limits larger than those of any burst reported in the first CHIME/FRB catalog,
which reported maximum values of $\tau_\mathrm{600\,MHz} = 0.09 \ \pm \ 0.006$~ms and fluence $\gtrsim$95~$\pm \ 41$~Jy~ms, respectively \citep{catalog1}.

One thing is immediately clear from Figure~\ref{fig:injections_comparison} --- in the fluence--scattering timescale--DM parameter space for bursts injected with a scattering timescale $>$100~ms at 600 MHz, there are not enough bursts in the injections dataset to fully sample the parameter space.
This lack of injected highly scattered bursts is likely due to the ``forward-modeling'' of determining which bursts to inject, necessary because the injections system lacked the technical capability to inject $5 \times 10^6$~synthetic pulses in a timely manner ($\S$4.1.1 in \citet{catalog1}).

While it is plausible that the high fluence of FRB~20200723B was sufficient to overcome the CHIME/FRB detection pipeline selection against highly scattered bursts, a more extensive injections campaign is necessary to better understand how real the fluence--scattering detection correlation may be, as well as what implications that may have for ``scattering horizons'' \citep[e.g.,][]{ocker+2022_horizons} for radio transient surveys.
To better understand selection biases, it is also necessary for the injections campaign to densely cover a larger parameter space of properties than what was observed.
FRB~20200723B also emphasizes the need for a better understanding of selection biases as correlated between burst properties.
Accordingly, a more thorough injections campaign is planned for the second CHIME/FRB catalog.

\section{Discussion \& Conclusion}
\label{sec:conclusion}

The most probable association of FRB 20200723B to a putative local Universe host galaxy, NGC~4602, has allowed us to explore its sightline properties.
We find that this FRB passes through the ``W-M sheet'', a filamentary structure associated with the Virgo Cluster \citep{kim+2016_virgocluster}.
Budgeting the total DM along the line of sight, we can constrain the average column density along the line of sight of this sheet to be $\Sigma~<~5.5 \times 10^{20}$~cm$^{-2}$, consistent with gas density results from IllustrisTNG.
We also combine this result with Compton $y$ values corresponding to filaments to obtain a lower limit on the average temperature in the W-M sheet, $T > 3 \times 10^5$~K.
While a filament is structurally distinct from a sheet, and thus this temperature limit is perhaps not the most accurate or constraining, one can imagine improved constraints in the near future for specific filaments and sheets with updated tSZ observations of sheets and better-localized FRBs in the local Universe.
We can also very conservatively constrain the average free electron density along the line of sight of this sheet to be $\langle n_e \rangle~<~4.6^{+9.6}_{-2.0} \times 10^{-5}$~cm$^{-3}$, where the main source of uncertainty is the redshift-independent distance measurement of NGC~4602.
This free electron density constraint is also in broad agreement with simulations.

We note this free electron density upper limit corresponds to a filamentary DM contribution of DM$_\mathrm{sheet}~\sim~179$~pc~cm$^{-3}$, over 70\% of the total observed DM of $\sim$244~pc~cm$^{-3}$ of FRB~20200723B.
While \citet{walker+2024} find in simulations that filamentary contributions to DM dominate starting from $z=0.1$ (the non-zero redshift snapshot they consider), it has not been standard to attribute so large a percentage of DM of FRBs to the intervening cosmic web.

With this DM budgeting, using formalism developed by \citet{cordes+2016}, \citet{ocker+2021_halos}, and \citet{cordes+2022_dmtau} for scattering from ionized cloudlets, we find that the large scattering timescale of FRB~20200723B is reasonably attributable to its host galaxy.
Adopting a model where there are cloudlets that could cause radio-wave scattering within the warm-hot intergalactic medium dominating galaxy cluster filaments,
we find that FRB~20200723B is unlikely to have its scattering due to the filamentary structure it resides within, as the physical configurations required are rather fine-tuned.
We also posit that this FRB, despite its high scattering timescale, is unlikely to originate from an unseen background galaxy behind the disk of NGC~4602, in large part due to chance alignment probabilities.
More precise localizations (e.g., sub-arcsecond instead of $\sim$arcminute) would provide clarity on such a scenario.

We compare the DM and scattering measurements of FRB~20200723B with those of pulsars compiled by \citet{ocker+2024_hii}, and find it physically plausible that if the scattering layer of FRB~20200723B is within its host galaxy,
it may be a result of intersecting an H\,\textsc{ii} region within NGC~4602 along its sightline.
Such an intersection probability is increased by the high inclination angle of NGC~4602.
The scattering excess compared to the mean scattering--DM relation is comparable to the scattering excess of pulsars found to be associated with foreground ionized structures;
many more pulsars have also been identified to have sightlines through foreground H\,\textsc{ii} regions \citep{ocker+2024_hii}.
Although rarer, scattering due to supernova remnants or eclipsing materials have also been observed for Milky Way pulsars at levels comparable to the scattering excess seen for FRB~20200723B.
We find a sightline intersection with an H\,\textsc{ii} region within the host galaxy to be a more natural explanation for the large scattering timescale of FRB~20200723B,
compared to requiring an improbable (though not impossible) alignment of physical parameters to explain the scattering if it were attributable to the W-M sheet (Section~\ref{subsec:scattering_wm_sheet}), or to a scenario where the unseen true host galaxy lies behind NGC~4602 (Section~\ref{subsec:scattering_background}).
We also note that the more DM we attribute to a screen within the host galaxy of NGC~4602, the less the observed scattering is in excess of the empirical mean scattering--DM relation, and also the less DM is attributable to the filamentary structure, thus making the free electron density constraints tighter for the W-M sheet.

In over five years of operation, CHIME/FRB has seen only two bursts with scattering timescales $\tau_\mathrm{400\,MHz} >$1~second: FRB~20191221A and FRB~20200723B.
That we even see FRBs with such high scattering timescales is surprising, given that \citet{catalog1} have noted that CHIME/FRB has an extreme observation selection bias against bursts with scattering timescales $\tau_\mathrm{600\,MHz} >$10~ms.
What both FRB 20200723B and FRB 20191221A have in common is high burst fluences, suggesting that the selection against large scattering can be overcome with sufficiently bright bursts.
However, the parameter space explored by injections for high fluence-high scattering bursts is sparse.
Thus, until a more comprehensive injections campaign is done,
there is no way to quantify whether these high fluence-high scattering bursts are intrinsically rare, or just observationally selected against by CHIME/FRB.

FRB~20200723B already demonstrates the ability of FRBs to sensitively probe ionized media along their sightlines using both their DM and their scattering properties.
We note that local Universe FRBs hold especially promising sightline constraints given the amount of detailed DM budgeting that it is possible to do.
Bursts such as FRB~20200723B, but with a full baseband capture for polarimetric information, can reveal even more information about the interplay between fluence, scattering timescale, and local magneto-ionospheric environment for FRBs.
As FRB~20200723B was detected before any CHIME/FRB~Outriggers stations came online, only an $\sim$arcminute-scale localization was possible;
with sub-arcsecond localizations, observations of FRB~20200723B-like sources could start to further probe possible CGM substructure of other galaxies \citep[e.g.,][]{faber+2024_scattered}.
With the rapidly growing sample of FRBs, studies of their propagation effects will be a powerful probe of the Universe.

\section*{}
\noindent
We acknowledge that CHIME is located on the traditional, ancestral, and unceded territory of the Syilx/Okanagan people. We are grateful to the staff of the Dominion Radio Astrophysical Observatory, which is operated by the National Research Council of Canada.  CHIME is funded by a grant from the Canada Foundation for Innovation (CFI) 2012 Leading Edge Fund (Project 31170) and by contributions from the provinces of British Columbia, Qu\'{e}bec and Ontario. The CHIME/FRB Project is funded by a grant from the CFI 2015 Innovation Fund (Project 33213) and by contributions from the provinces of British Columbia and Qu\'{e}bec, and by the Dunlap Institute for Astronomy and Astrophysics at the University of Toronto. Additional support was provided by the Canadian Institute for Advanced Research (CIFAR), the Trottier Space Institute at McGill University, and the University of British Columbia.

The authors would like to thank Jim Cordes for kindly providing pulsar data.

\allacks

\facility{CHIME}

\software{
    \texttt{APLpy} \citep{aplpy2012, aplpy2019},
    \texttt{astropath} \citep{aggarwal+2021_PATH},
    \texttt{astropy} \citep{astropy:2013, astropy:2018, astropy:2022},
    \texttt{fitburst} \citep{fonseca+2023_fitburst},
    \texttt{matplotlib} \citep{matplotlib},
    \texttt{NE2001p} \citep{ocker_cordes_2024_NE2001p},
    \texttt{numpy} \citep{numpy},
    \texttt{PyGEDM} \citep{pygedm},
    \texttt{reproject} \citep{reproject2020},
    \texttt{RM-CLEAN} \citep{heald+2009_rmclean},
    \texttt{RM-synthesis} \citep{brentjens_deBruyn_2005_RM_synth}, 
    \texttt{RM-tools} \citep{RMTools},
    \texttt{scipy} \citep{scipy}
}

\bibliographystyle{aasjournal}
\bibliography{refs}

\end{document}